\documentclass{aa}
\usepackage{graphics}
\begin{document}

\newcommand{\alf}{$\alpha$}
\newcommand{\smu}{$\sigma_{\mu}$}
\newcommand{\muacd}{$\mu_{\alpha}$.cos$(\delta)$}
\newcommand{\mud}{$\mu_{\delta}$}
\newcommand{\mulcb}{$\mu_{l}$.cos$(b)$}
\newcommand{\mub}{$\mu_{b}$}
    
\thesaurus{(03.13.5;04.03.1;05.01.1;08.06.2;08.11.1)}

\title{Proper motions of pre-main sequence stars in southern star-forming 
regions
       \thanks{Based on observations made at {\sc Valinhos} CCD
       Meridian Circle. Based on measurements made with {\sc MAMA} automatic
       measuring machine}
       \thanks{Table \ref{Tab4} will be only available in electronic form at the CDS via
       anonymous ftp to cdsarc.u-strasbg.fr (130.79.128.5) or via 
       http://cdsweb.u-strasbg.fr/Abstract.html}
      }

\author{R.~Teixeira \inst{1,2}
        \and
        C.~Ducourant \inst{2,1}
        \and
        M.J. Sartori \inst{1}
        \and
        J.I.B. Camargo \inst{1}
        \and
        J.P.~P\'eri\'e \inst{2}
        \and
        J.R.D. L\'epine \inst{1}
        \and
        P. Benevides-Soares \inst{1}}

\institute{Departamento de Astronomia, Universidade de S\~ao Paulo, Av.
           Miguel St\'efano 4200, 04301-904 S\~ao Paulo, Brazil.
           \and
           Observatoire de Bordeaux, UMR 5804, CNRS/INSU, BP 89,
           F-33270 Floirac, France.}

\offprints{teixeira@orion.iagusp.usp.br}
\date{Received 30 March 2000 / Accepted 15 June 2000}
\maketitle
\markboth{Teixeira et al.: Proper motions of pre-main sequence stars in southern 
star forming regions}{}

\begin{abstract}
%$$ $$
We present proper motion measurements of pre-main sequence (PMS) stars 
associated with  major star-forming regions of the
southern hemisphere (Chamaeleon, Lupus, Upper Scorpius - Ophiuchus, Corona 
Australis), 
situated in the
galactic longitude range $l = 290\degr $ to $l = 360\degr$. A list of
PMS stars as complete as possible was established based on the Herbig and
Bell catalogue and many new catalogues like the PDS survey, the catalogue of
Herbig Ae/Be stars by Th\'e et al. (\cite{the}), X-rays surveys, etc.
The measurements made use of public material (mainly
{\sc AC2000} and {\sc USNO--A2.0} catalogues) as well as
scans of {\sc SERC--J} Schmidt plates with the {\sc MAMA} measuring 
machine
(Paris) and Valinhos CCD meridian circle observations (Brazil).
We derived proper motions for 213 stars, with an accuracy
of 5 to 10 mas/yr depending mainly on the difference of epochs between the 
position sources.
The main characteristics of the sample are discussed. We show that
systematic motions of groups of stars exist, which are not explained by
the reflex solar motion.

\end{abstract}

\section{Introduction}
%---------------------
Analysis of the motion of pre-main sequence (PMS) stars and of
related groups of young stars provide essential tests of star
formation models. Different space velocities and velocity gradients
of the stellar associations can be derived from the major star formation
scenarios, like sequential star formation, star formation by high-velocity
clouds, Gould's Belt models, etc. Proper motion measurements
of the members of these associations provide one way to discriminate among these
predictions. The PMS stars are supposed to be  sufficiently young to be
very close to their birthplaces and to have velocities still very similar
to the initial ones, so that one can get clear constraints  on the birth
mechanism.

In this work, we investigated the PMS stars of an extended region of aligned
molecular clouds and OB associations that includes the Chamaeleon, Lupus, Upper 
Scorpius - Ophiuchus
and Corona Australis regions. This selected area is specially interesting
because the associations  are close enough to the Sun (100-150 pc),
so that a refined kinematical study can be made.

The HIPPARCOS mission (ESA 1997) provided  accurate measurements of positions,
parallaxes and proper motions, for OB stars brighter than V=10 mag, allowing
to study the kinematics of the same regions (de Zeeuw et al. \cite{zeu}). Proper 
motions
of PMS stars associated with these star-forming complexes,
based on HIPPARCOS data, were recently obtained (Frink et al. \cite{fri}, 
Neuh\"auser \&
Brandner \cite{neu}, Wichmann et al. \cite{wib}, Bertout et al. \cite{ber}), 
providing a first
comparison of space velocities of different groups of stars. However,
since most PMS stars are fainter than the limiting magnitude of
HIPPARCOS, the total number of measured PMS stars is still small.

In this work, we present proper motion determination for 213 PMS stars as
faint as V$\sim$16 mag lying in the galactic longitude range $l = 290\degr$ 
to $l =  360\degr$.

In the following sections, we present the method used to derive
proper motions based on different combinations of first and second epoch
measurements, we discuss the quality of our results and present
a first analysis of this large sample of proper motions obtained for the PMS 
stars.

\section{Data}
%-------------
For this work, we collected  a list of well known and candidate T Tauri and 
Herbig Ae/Be (hereinafter HAeBe)
stars as exhaustive as possible for the
studied regions in the literature (Alcal\'a \cite{ala}; Alcal\'a 
et al. \cite{alb}; Bertout et al. \cite{ber}; Brandner et al. \cite{bra}; 
Casanova
 et al. \cite{cas}; Feigelson \& Kriss \cite{fea}; Feigelson
et al. \cite{feb}; Gauvin \& Strom \cite{gau}; Gregorio-Hetem et al. \cite{gre}; 
Hartigan \cite{har}; Herbig \& Bell \cite{heb}; Krautter et al.
\cite{kra}; Malfait et al. \cite{mal}; Marraco \& Rydgren \cite{mao}; 
Mart\'{\i}n et al. \cite{mat} , Preibisch et al. 
\cite{pre}; Schwartz \cite{sca}; Th\'e et al. \cite{the}; Torres et al. 
\cite{toa}; van den Ancker et al. \cite{vaa}; 
van den Ancker et al. \cite{vab}; Walter et al. \cite{wab}; Wichmann et al. 
\cite{wia}; Wilking et al. \cite{wil}). 
With these data, we constructed an input catalogue containing about 680 stars 
with 
their approximate positions and magnitudes. In this set of stars, we could 
identify and 
measure proper motions of 213 PMS stars as faint as V$\sim$16 mag. We 
also present, separately, measurements for 29 stars that were previously 
considered as PMS, but subsequently classified differently by Covino et al. 
(\cite{cov}) and Wichmann et al. (\cite{wic}), who did not confirm their PMS 
nature. 
These authors obtained high resolution spectra, and compared the Li line 
equivalent widths to those of the Pleiades stars, considered by them as a 
frontier 
between PMS and non-PMS stars.

The proper motions were obtained from the comparison
between several sources including current epoch CCD meridian observations
performed at Valinhos Observatory (Viateau et al. \cite{via}) and old
{\sc SERC--J} Schmidt plates measured {with} {\sc MAMA}
automatic measuring machine (Guibert et al. \cite{gui}). These data were 
combined with
published positions extracted from the following catalogues:
AC2000 (Urban et al. \cite{ura}), {\sc USNO--A2.0} (Monet \cite{mon}),
HIPPARCOS and Tycho (ESA \cite{esa}). One star had its proper motion obtained
from a combination of CCD meridian observation and a Digitized Sky 
Survey\footnote{The Digitized Sky Survey was 
produced at the Space Telescope Science Institute under US Government grant NAG 
W--2166.} (DSS)
image.

\subsection{CCD meridian observations}
%-------------------------------------
The most recent observations used here were obtained with the Valinhos CCD 
meridian circle (Dominici et al. \cite{dom})
in 1998 and 1999.  This instrument is installed in Valinhos at the Abrah\~ao de 
Moraes Observatory
(Latitude $-23\degr 00\arcmin 06\arcsec$, Longitude $+46\degr 58\arcmin 
03\arcsec$)
which belongs to the S\~ao Paulo University -- Brazil.

The CCD detector has 512x512 square pixels of 19$\mu m$
( 1 pixel = 1.5$^{\prime\prime}$ square) and works in drift scan mode. In this 
mode,
the telescope is fixed and the electric charges are moved along the columns of 
the
CCD with the same velocity as the transit, which depends on the declination. For
$\delta = 0\degr$ the integration time interval is 51 seconds. The observed 
field has
therefore an arbitrary length in right ascension (typically one hour) and is 
13$^{\prime}$ wide in declination. 
The magnitude limit is about V=16.0 mag. The astrometric and photometric 
observational precisions
depend  on the magnitude and in the best interval ($9.0<$V$<14.0$) are, 
respectively,
approximately $0.050\arcsec$ in both coordinates and 0.05 magnitudes. 
At the detection limit, positional measurements are less accurate 
and the mean square 
error may reach $0.100\arcsec$. Observations are carried out with a filter 
CG495$+$BG38  
(bandpass from 5200 to 6800 \AA), which is wider than Johnson filter. 
Notwithstanding, the resulting magnitudes are close to 
the visual standard magnitude system (Dominici et al. \cite{dom}).
The final positions are obtained by a global reduction procedure 
(Benevides-Soares
\& Teixeira \cite{ben}; Teixeira et al. \cite{tei}) using the ACT (Urban et al. 
\cite{urb}) as the reference catalogue. The data treatment is made by 
means of a software package developed and maintained by J.F. Le Campion 
(Bordeaux Observatory).

In this work, we observed seven strips in the Chamaeleon, Lupus and Upper 
Scorpius - Ophiuchus 
regions containing many PMS stars identified in the literature. The length in 
R.A. of the fields 
presented in Table \ref{Tab1} was defined
to ensure a minimum of 20 reference stars in each strip, necessary for an 
accurate
astrometric reduction, and their coordinates were defined to have the largest 
number
of PMS stars. In this table, we also present the central declination
and the number of observations of each strip. 

Each star considered here was observed at least three times.

\begin{table}[ht]
\caption{\label{Tab1}Valinhos CCD meridian observations strips (1998-1999)}
\begin{tabular}{cccc}
\hline\\[-3pt]
$\alpha_{min} $ & $\alpha_{max} $ &       $ \delta $     & Nobs.\\
   $[$h m$]$     &       $[$h m$]$     & $[\degr\;\;\arcmin]$ & \omit\\
\hline\\[-3pt]
15 38 & 16 25 &  $-$39 06 & 9 \\
15 39 & 16 18 &  $-$38 56 & 6 \\
16 11 & 16 51 &  $-$24 04 & 3 \\
15 50 & 17 20 &  $-$24 12 & 6 \\
10 39 & 11 27 &  $-$76 39 & 3 \\
10 39 & 11 27 &  $-$76 25 & 3 \\
12 43 & 13 19 &  $-$76 39 & 8 \\
\hline
\end{tabular}
\end{table}

\subsection{Schmidt plate material}
%----------------------------------
Thirteen {\sc SERC--J} $6.5\degr \times 6.5 \degr$ Schmidt plates have 
been
digitized at the {\sc MAMA} measuring machine (Guibert et al. \cite{gui}),
which provides at the present time the most accurate measurements
(repeatability of $0.4\mu m$). For each plate, a catalogue of ($x$,$y$), flux 
and area
has been produced for about 1\,000\,000 objects detected.

Each plate was then astrometrically reduced from ($x$,~$y$) to
($\alpha$,~$\delta$) with reference stars from the ACT catalogue
(Urban et al. \cite{urb}). The mean residual of these reductions was about
0.25$\arcsec$ in both coordinates. We give in Table
\ref{Tab2} the
list of the {\sc SERC--J} plates used in this work to derive proper motions
for the known PMS stars. In this table are shown the numbers of the plates,
their observation epochs and their central coordinates.

\begin{table}[ht]
\caption{\label{Tab2}Schmidt {\sc SERC--J} plates measured}
\begin{tabular}{cccl}
\hline\\[-3pt]
Number    & Epoch & $\alpha$ [{\rm J}2000] & \hskip 6pt $\delta$ [{\rm J}2000]\\
\hline\\[-3pt]
038   &  1976.255&  11 00 53& $-$75 17 43\\
039   &  1976.252&  12 08 42& $-$75 16 42\\
040   &  1980.531&  13 15 53& $-$75 15 50\\
387   &  1975.268&  15 15 09& $-$35 11 07\\
388   &  1974.474&  15 39 15& $-$35 11 51\\
329   &  1978.331&  15 39 19& $-$40 09 43\\
330   &  1975.422&  16 05 22& $-$40 08 07\\
517   &  1978.578&  16 33 03& $-$25 06 17\\
519   &  1976.411&  17 16 57& $-$25 02 46\\
521   &  1974.558&  18 01 04& $-$25 00 02\\
396   &  1974.460&  18 51 23& $-$34 58 16\\
337   &  1974.561&  19 07 27& $-$39 55 16\\
397   &  1977.528&  19 15 16& $-$34 54 52\\
\hline
\end{tabular}
\end{table}

\begin{figure*}
\rotatebox{-90}{\resizebox{12cm}{!}{\includegraphics{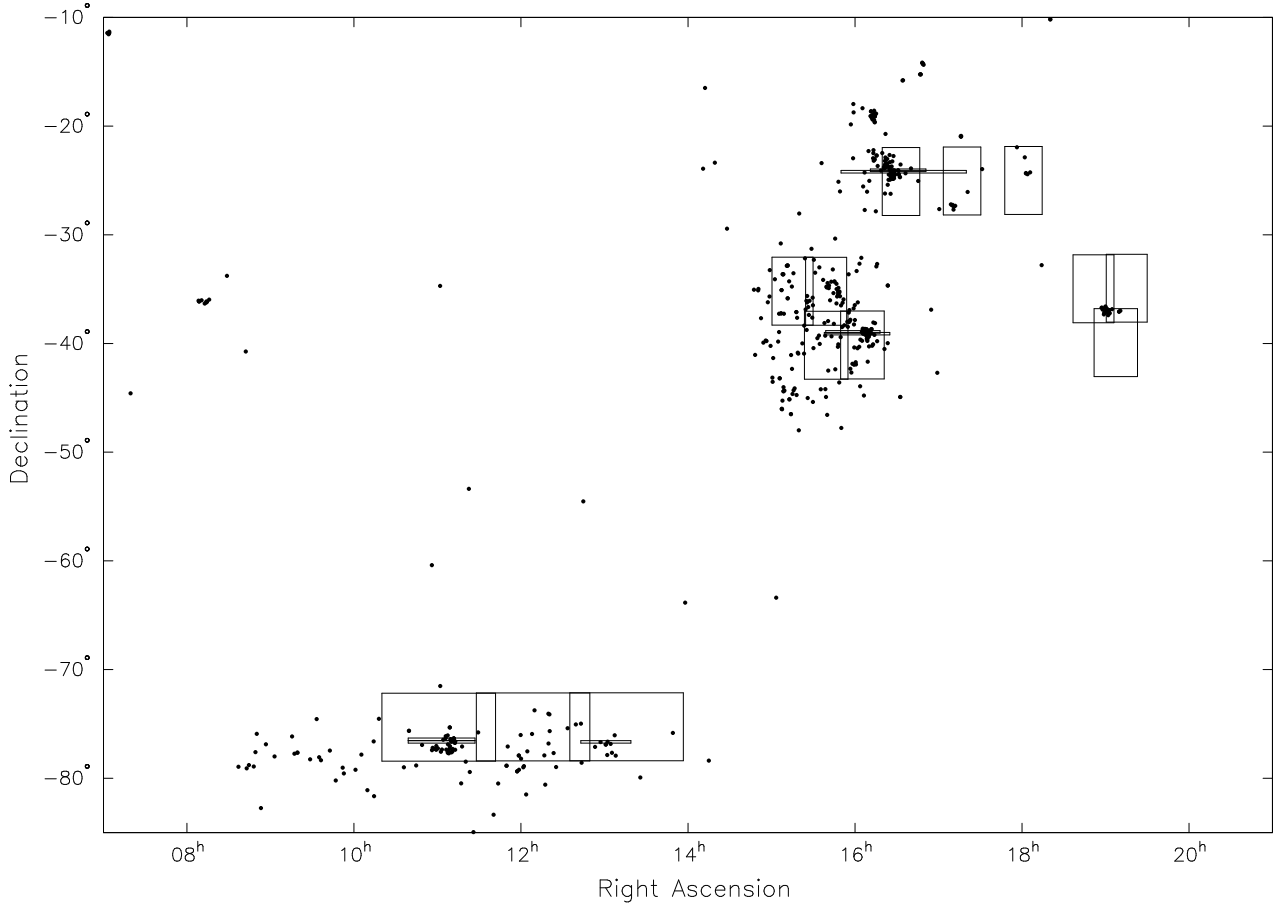}}}
\caption{Survey area. Rectangles = Schmidt plates, Strips = transit circle frames}
\label{Fig1}
\end{figure*}

The Schmidt plates and Valinhos strips distribution in the studied region of the 
sky are given
in Fig. \ref{Fig1}, along with the searched PMS stars from our input catalogue.

\subsection{Other data}
%----------------------
For a good determination of proper motions, observations from well separate
epochs are needed. Unfortunately, the catalogue with the oldest mean epoch
(AC2000, Urban et al. \cite{ura}), gives positions for stars  essentially 
brighter 
than
V=12 mag. We included these positions in our calculations, when available.
For fainter stars (V$>$12), proper motions have been determined 
over a shorter time interval and so with a degraded precision. One of these
stars, Sz 108, although observed at the Valinhos observatory, could not be found 
in the other
catalogues. Its position was determined from the DSS (plate identification: 
02F8, region: S330), 
using the IRAF software packages.  

We also included many other available astrometric data as from {\sc USNO--A2.0}
Catalogue (Monet \cite{mon}), Tycho and HIPPARCOS positions (ESA \cite{esa}) to 
better constrain the proper 
motion determination and to provide a larger number of proper motions.

\begin{table}[ht]
\caption{\label{Tab3}Evaluation of the precision of the proper motions}
\begin{tabular}{lccccc}
\hline\\[-3pt]
Sources    & $\Delta$t & $ t_{0}$ &  $\sigma_{\alpha_{0}}$ & \smu & N \\
\omit      & [yr] & [yr] & [$\arcsec$] & [mas/yr] & \\
\hline\\[-3pt]
AEU     &75  &1948.33 &   0.144&   4  &    46 \\
AU      &70  &1935.00 &   0.177&   5  &    48 \\
ATU     &91  &1989.68 &   0.030&   3  &    42 \\
EUV     &28  &1996.11 &   0.048&   7  &    25 \\
AHTU    &91  &1991.25 &   0.001&   3  &    20 \\
AETU    &91  &1989.47 &   0.029&   3  &    22 \\
AHT     &91  &1991.25 &   0.001&   3  &     7 \\
AEHTU   &91  &1991.25 &   0.001&   3  &     7 \\
AEUV    &98  &1992.68 &   0.047&   2  &     3 \\
AE      &75  &1937.50 &   0.177&   5  &     3 \\
AEHT    &91  &1991.25 &   0.001&   3  &     2 \\
AEHTUV  &98  &1991.25 &   0.001&   2  &     2 \\
UV      &28  &1996.92 &   0.049&   9  &     2 \\
AHU     &81  &1991.25 &   0.001&   3  &     2 \\
AEH     &78  &1991.25 &   0.001&   3  &     2 \\
AH      &85  &1991.25 &   0.001&   3  &     2 \\
AHTV    &98  &1991.25 &   0.001&   3  &     1 \\
AHTUV   &98  &1991.25 &   0.001&   3  &     1 \\
AETUV   &98  &1991.66 &   0.025&   2  &     1 \\
ATUV    &98  &1991.83 &   0.025&   3  &     1 \\
AUV     &98  &1996.23 &   0.025&   3  &     1 \\
AT      &91  &1989.31 &   0.030&   3  &     1 \\
DV      &23  &1997.98 &   0.050&  17  &     1 \\
\hline
\end{tabular}

In 1$^{\rm st}$ column, A = AC2000, D = DSS, E = SERC--J, H = HIPPARCOS, 
T = Tycho, U = USNO--A2.0, V = Valinhos.
\end{table}

\section{Proper motion determination}
%-------------------------------------
From our initial sample of about 680 stars, we could measure accurate
proper motions for 242 of them. These stars have magnitudes within the 
range 6$<$V$<$16.
Proper motions have been determined only when the time basis was longer
than 20 years. In most cases this time basis was longer than 50
years, reaching in some individual cases more than 100 years. About 30 
proper motions were determined with a time basis shorter
than 50 years. The mean time basis was 80 years.

For the remaining stars, the main reasons of their absence in our final 
catalogue
are that either their magnitudes laid outside the magnitude range covered by our 
data sources
or the request of a time basis longer than 20 years was not fulfilled.

The proper motion calculation was performed in the usual way, via a weighted 
least
squares method (Eqs. 1 to 5).

\begin{eqnarray}
    t_{0}                   & = &\frac{\sum p_{i}t_{i}}{\sum p_{i}}\\
    \alpha_{0}              & = &\frac{\sum \alpha_{i}p_{i}}{\sum p_{i}}\\ 
    \mu_{\alpha}            & = &\frac{\sum p_{i}\alpha_{i}(t_{i}-t_{0})}
{\sum p_{i}(t_{i}-t_{0})^{2}}\\
    \sigma_{\alpha_{0}}^{2} & = &\frac{1}{\sum p_{i}}\\
    \sigma_{\mu_{\alpha}}^{2}        & = &\frac{1}{\sum p_{i}(t_{i}-t_{0})^{2}}
\end{eqnarray}\noindent
where $p_{i}=\frac{1}{\sigma_{i}^{2}}$, and $t_{i}$ is the epoch of 
the position
for a given star $i$. The same calculation holds for the declination.

We have assumed the following precisions for the various data : $\sigma =
0.25\arcsec$
for AC2000, USNO--A2.0 and {\sc SERC--J} positions, $\sigma = 0.001 
\arcsec$ for HIPPARCOS,
$\sigma = 0.030 \arcsec$ for Tycho and $\sigma = 0.050\arcsec$ for Valinhos 
positions.

We present the precision of the derived proper motions with various 
material in Table \ref{Tab3}, where $\Delta$t is the mean time basis for the 
groups of sources given in column one, and N stands for the number of stars 
whose proper motions were obtained from the combination of these sources. 

We give in Tables \ref{Tab4} and \ref{Tab5}
the derived mean positions and proper motions for the 213 PMS stars. 
In these tables, the objects
were separated in T Tauri stars (Table \ref{Tab4}) and HAeBe stars (Table 
\ref{Tab5}). Table \ref{Tab6}
lists the 29 non-PMS ROSAT stars in Chamaeleon and Lupus (see section 2). In 
addition, a traditional separation by region was adopted. 

In the second column of these tables, magnitudes are taken
preferably from Tycho catalogue ($V_{T}$). For the Valinhos non-Tycho stars, 
Valinhos visual magnitudes are used. Otherwise,
magnitudes are taken from the literature, as indicated by the references (column 
3) explicited at the end of Table \ref{Tab6}. For stars 
RX J1621.4--2332ab and RX J25.3--2402 we give the B magnitude from the AC2000. 
Magnitudes taken from reference $[10]$ are also B magnitudes.

\begin{figure*}
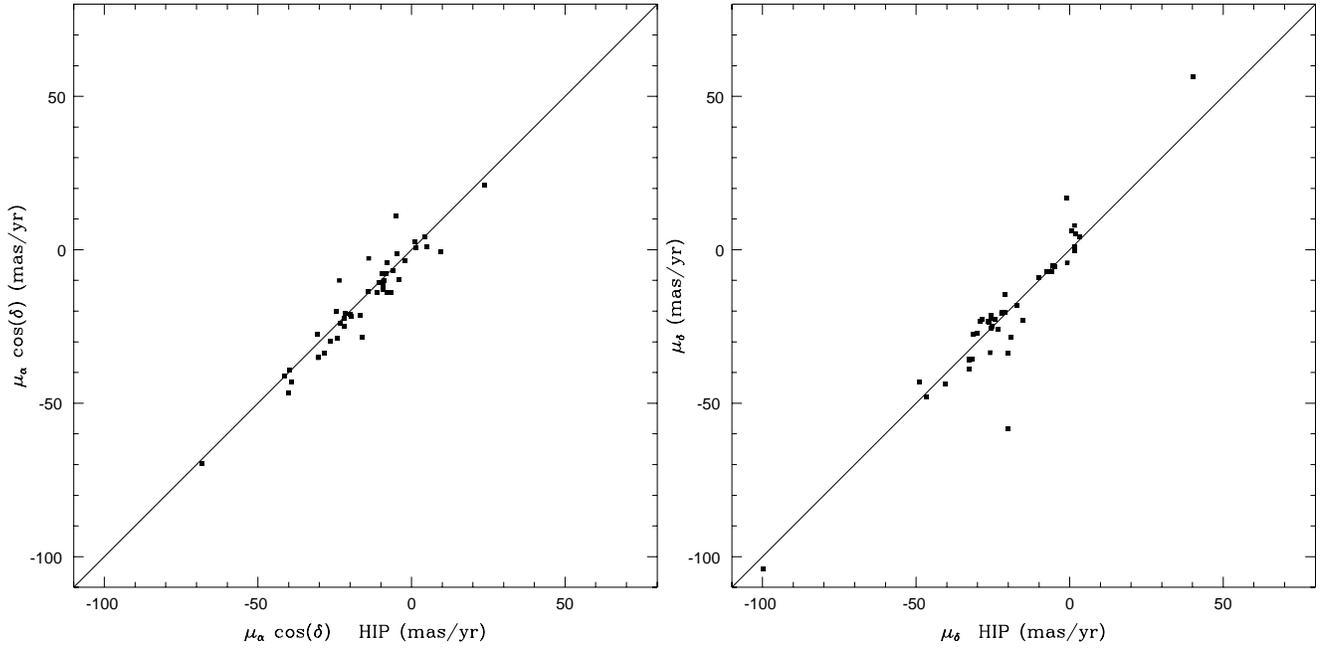

\resizebox{17.5cm}{!}{\includegraphics{ms9789.f2a}\includegraphics{ms9789.f2b}}
\caption{Comparison of our proper motions in right ascension (left 
panel) and declination (right panel) with HIPPARCOS ones.}
\label{Fig2}
\end{figure*}

\begin{figure*}
\resizebox{17.5cm}{!}{\includegraphics{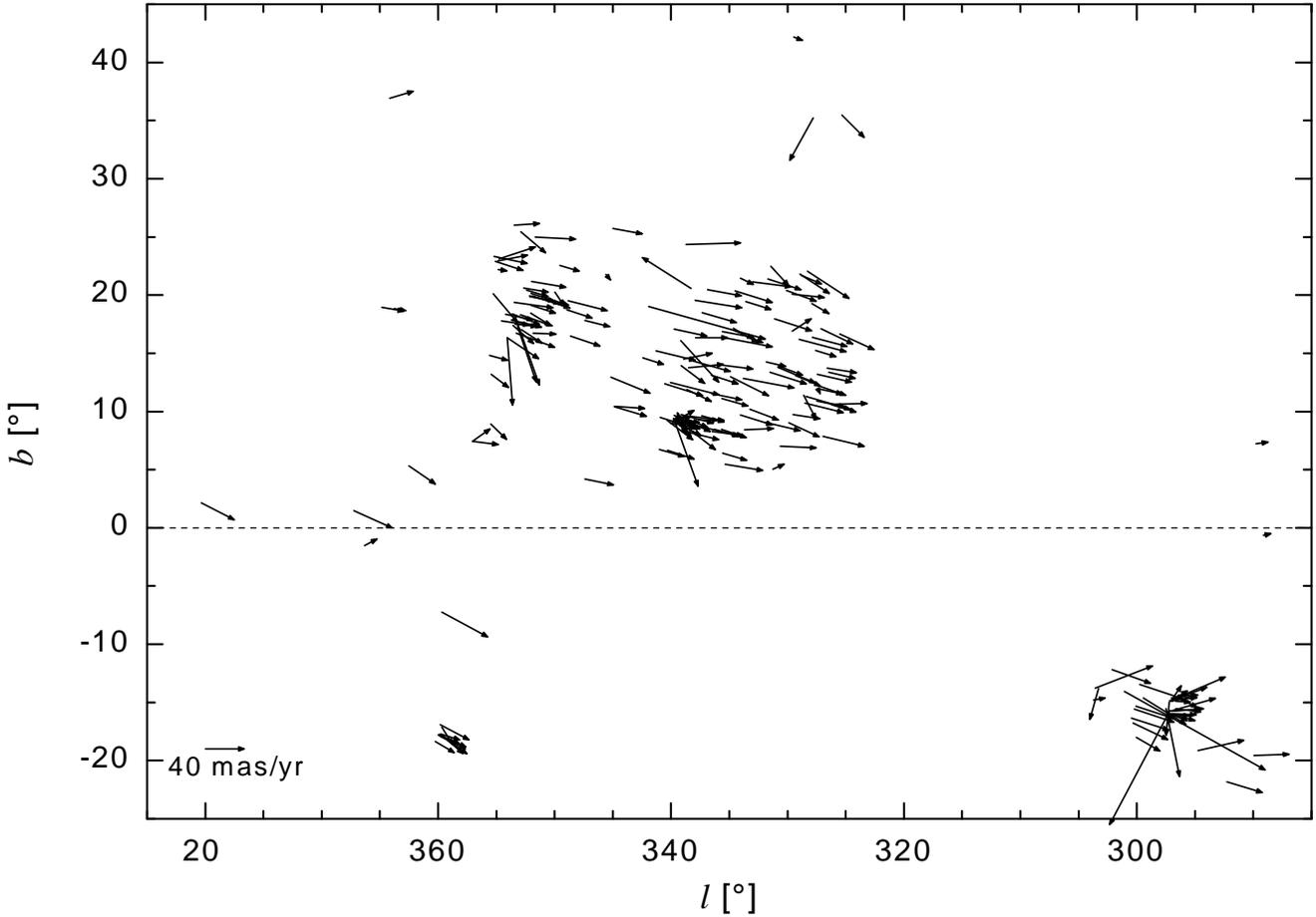}}
\caption{Positions and proper motions of PMS stars depicted in 
Figs. \ref{Fig4} to \ref{Fig7} -- upper panels.}
\label{Fig3}
\end{figure*}

\begin{table*}[ht]
\scriptsize
{
\caption{\label{Tab4}Proper motions for the T Tauri stars}
\begin{tabular}{lccccccccccc}
\hline\\[-2.8pt]
Name              &  Mag & Ref. & Sources & Epoch    & 
\multispan2{\hfill{$\alpha$\hbox{\hskip 15pt} [J2000]\hbox{\hskip 15 pt} 
$\delta$}\hfill} & 
$\mu_{\alpha}{\rm cos}\delta$ & $\mu_{\delta}$ & $\mu_{l}{\rm cos}b$ & $\mu_{b}$ 
& Other PM's\\[2.8pt] 
\omit             & \omit& \omit& \omit   &\omit     & 
\multispan2{\hfill{$\overline{{\rm{[h\hskip 6pt m\hskip 6pt s]}}\hbox{\hskip 
50pt}\hbox{\hskip 
-12pt}[^{{\rm o}}\;\;\;\arcmin\;\;\;\arcsec]}$}\hfill} & 
\multispan4{\hfill{$\overline{\hbox{\hskip 60pt}{\rm [mas/yr]}\hbox{\hskip 
60pt}}$}\hfill} & 
\omit\\[2.8pt] 
\hline\\[-7.8pt]
Chamaeleon\\[2.8pt]
\hline
RXJ0837.0-7856    & 10.7 &  [T] & ATU	 & 1988.731 & 08 36 56.352  &$-$78 56 
45.95 & \hskip 3.5pt$-$29  & \hskip 3.5pt$+$26  & \hskip 3.5pt$-$37  & \hskip 
3.5pt$-$11  &  1,4 \\[2.8pt]
RXJ0850.1-7554    & 10.6 &  [T] & ATU	 & 1988.580 & 08 50 05.502  &$-$75 54 
38.55 & \hskip 3.5pt$-$20  & \hskip 3.5pt$+$31  & \hskip 3.5pt$-$37  & \hskip 
7.0pt$+$2	&  1,4 \\[2.8pt]
HD 86356          & 10.2 &  [T] & ATU	 & 1988.957 & 09 51 50.784  &$-$79 01 
38.25 & \hskip 3.5pt$-$28  & \hskip 3.5pt$+$40  & \hskip 3.5pt$-$48  & \hskip 
3.5pt$+$11  & 1,3,4\\[2.8pt]
Sz 6              & 11.5 &  [H] & AEH	 & 1991.249 & 10 59 07.031  &$-$77 01 
40.33 & \hskip 3.5pt$-$25  & \hskip 7.0pt$+$8   & \hskip 3.5pt$-$26  & \hskip 
7.0pt$-$4	& 1,2,4\\[2.8pt]
CHXR8             & 11.1 &  [T] & AETU   & 1988.893 & 11 00 14.581  &$-$77 14 
38.11 & \hskip 3.5pt$-$25  & \hskip 3.5pt$+$13  & \hskip 3.5pt$-$28  & \hskip 
7.0pt$+$1	&   1  \\[2.8pt]
CS Cha            & 11.6 & [12] & AEU	 & 1961.796 & 11 02 25.116  &$-$77 33 
35.95 & \hskip 3.5pt$-$24  & \hskip 3.5pt$+$10  & \hskip 3.5pt$-$26  & \hskip 
7.0pt$-$1	&   4  \\[2.8pt]
CHXR11            & 11.5 &  [8] & AEU	 & 1961.796 & 11 03 11.760  &$-$77 21 
04.68 & \hskip 3.5pt$-$29  & \hskip 3.5pt$+$16  & \hskip 3.5pt$-$33  & \hskip 
7.0pt$+$3	&   4  \\[2.8pt]
CT Cha            & 12.5 &  [V] & EUV	 & 1998.444 & 11 04 09.131  &$-$76 27 
19.30 & \hskip 3.5pt$-$15  & \hskip 7.0pt$+$5   & \hskip 3.5pt$-$15  & \hskip 
7.0pt$-$1	&      \\[2.8pt]
CED 110           & 11.2 &  [4] & AE	 & 1944.719 & 11 06 16.188  &$-$77 22 
08.94 & \hskip 3.5pt$+$25  &	       $-$370 & 	    $+$175 &		
 $-$328 &      \\[2.8pt]
DI Cha            & 11.0 &  [T] & AHTU   & 1991.249 & 11 07 20.788  &$-$77 38 
07.32 & \hskip 3.5pt$-$29  & \hskip 7.0pt$+$4   & \hskip 3.5pt$-$28  & \hskip 
7.0pt$-$8	& 1,2,4\\[2.8pt]
VW Cha            & 12.9 &  [4] & AEU	 & 1961.796 & 11 08 01.806  &$-$77 42 
28.75 & \hskip 3.5pt$-$23  & \hskip 7.0pt$+$7   & \hskip 3.5pt$-$24  & \hskip 
7.0pt$-$3	&   4  \\[2.8pt]
Glass\#Ia,b       & 12.8 & [14] & AEU	 & 1961.796 & 11 08 15.411  &$-$77 33 
53.45 & \hskip 3.5pt$-$40  & \hskip 3.5pt$+$33  & \hskip 3.5pt$-$50  & \hskip 
3.5pt$+$14  &   4  \\[2.8pt]
VZ Cha            & 13.2 &  [V] & EUV	 & 1998.444 & 11 09 23.802  &$-$76 23 
20.73 & \hskip 3.5pt$-$11  & \hskip 3.5pt$+$14  & \hskip 3.5pt$-$15  & \hskip 
7.0pt$+$9	&      \\[2.8pt]
WY Cha            & 13.9 &  [V] & EUV	 & 1998.444 & 11 10 07.093  &$-$76 29 
37.62 & \hskip 7.0pt$-$3   & \hskip 3.5pt$+$18  & \hskip 3.5pt$-$10  & \hskip 
3.5pt$+$15  &      \\[2.8pt]
WZ Cha            & 15.5 &  [V] & EUV	 & 1998.444 & 11 10 53.274  &$-$76 34 
31.33 & \hskip 3.5pt$-$29  & \hskip 3.5pt$+$28  & \hskip 3.5pt$-$37  & \hskip 
3.5pt$+$14  &      \\[2.8pt]
CHX 18N           & 12.0 &  [V] & EUV	 & 1998.444 & 11 11 46.331  &$-$76 20 
08.82 & \hskip 3.5pt$-$23  & \hskip 3.5pt$+$13  & \hskip 3.5pt$-$27  & \hskip 
7.0pt$+$3	&      \\[2.8pt]
Sz 41             & 12.2 &  [V] & AEUV   & 1995.907 & 11 12 24.491  &$-$76 37 
05.88 & \hskip 3.5pt$-$25  & \hskip 3.5pt$+$14  & \hskip 3.5pt$-$29  & \hskip 
7.0pt$+$3	&   4  \\[2.8pt]
CV Cha            & 11.0 &  [T] & AEHTUV & 1991.252 & 11 12 27.759  &$-$76 44 
22.27 & \hskip 3.5pt$-$10  & \hskip 3.5pt$-$34  & \hskip 7.0pt$+$4   & \hskip 
3.5pt$-$35  & 1,2,4\\[2.8pt]
HM Anon           & 11.1 &  [4] & AEU	 & 1961.796 & 11 12 42.758  &$-$77 22 
00.26 & \hskip 3.5pt$-$40  & \hskip 3.5pt$-$60  & \hskip 3.5pt$-$14  & \hskip 
3.5pt$-$71  &   4  \\[2.8pt]
CHX 20E           & 12.2 &  [V] & EUV	 & 1998.444 & 11 12 43.030  &$-$76 37 
04.62 & \hskip 3.5pt$-$44  & \hskip 3.5pt$+$45  & \hskip 3.5pt$-$58  & \hskip 
3.5pt$+$25  &      \\[2.8pt]
RXJ1150.4-7704    & 12.0 &  [1] & AEU	 & 1950.578 & 11 50 28.926  &$-$77 04 
37.74 & \hskip 3.5pt$-$50  & \hskip 3.5pt$-$16  & \hskip 3.5pt$-$44  & \hskip 
3.5pt$-$27  &  4,6 \\[2.8pt]
T Cha             & 12.0 &  [H] & AH	 & 1991.248 & 11 57 13.656  &$-$79 21 
31.45 & \hskip 3.5pt$-$39  & \hskip 7.0pt$-$9   & \hskip 3.5pt$-$36  & \hskip 
3.5pt$-$17  & 1,2,4\\[2.8pt]
RXJ1158.5-7754a   & 10.8 &  [T] & AEHTU  & 1991.248 & 11 58 28.273  &$-$77 54 
29.61 & \hskip 3.5pt$-$41  & \hskip 7.0pt$-$4   & \hskip 3.5pt$-$39  & \hskip 
3.5pt$-$13  & 1,2,4\\[2.8pt]
RXJ1159.7-7601    & 11.1 &  [T] & AEHTU  & 1991.248 & 11 59 42.370  &$-$76 01 
26.13 & \hskip 3.5pt$-$47  & \hskip 7.0pt$-$6   & \hskip 3.5pt$-$45  & \hskip 
3.5pt$-$15  &  1,2 \\[2.8pt]
HD 104467         &  8.7 &  [T] & ATU	 & 1988.673 & 12 01 39.285  &$-$78 59 
16.87 & \hskip 3.5pt$-$40  & \hskip 7.0pt$-$4   & \hskip 3.5pt$-$39  & \hskip 
3.5pt$-$12  & 1,3,4\\[2.8pt]
RXJ1239.4-7502    & 10.4 &  [T] & AETU   & 1988.507 & 12 39 21.388  &$-$75 02 
39.06 & \hskip 3.5pt$-$40  & \hskip 3.5pt$-$12  & \hskip 3.5pt$-$40  & \hskip 
3.5pt$-$14  &  1,4 \\[2.8pt]
PDS 64            & 15.0 &  [V] & EUV	 & 1998.628 & 12 57 11.779  &$-$76 40 
11.50 & \hskip 3.5pt$+$10  & \hskip 3.5pt$-$32  & \hskip 7.0pt$+$9   & \hskip 
3.5pt$-$32  &      \\[2.8pt]
CM Cha            & 13.7 &  [V] & EUV	 & 1998.628 & 13 02 13.566  &$-$76 37 
57.77 & \hskip 3.5pt$-$60  & \hskip 3.5pt$+$21  & \hskip 3.5pt$-$60  & \hskip 
3.5pt$+$23  &      \\[2.8pt]
BF Cha            & 12.5 &  [6] & AEU	 & 1952.361 & 13 05 20.845  &$-$77 39 
01.32 & \hskip 3.5pt$-$12  & \hskip 7.0pt$+$2   & \hskip 3.5pt$-$12  & \hskip 
7.0pt$+$2	&   4  \\[2.8pt]
\hline
Lupus\\[2.8pt]
\hline
RXJ1410.8-2355    & 11.7 & [17] & AU	 & 1948.692 & 14 10 49.761  &$-$23 55 
27.16 & \hskip 3.5pt$-$13  & \hskip 3.5pt$-$30  & \hskip 3.5pt$-$23  & \hskip 
3.5pt$-$23  &      \\[2.8pt]
RXJ1412.2-1629    & 11.0 &  [T] & ATU	 & 1988.321 & 14 12 14.044  &$-$16 29 
53.31 & \hskip 7.0pt$-$7   & \hskip 7.0pt$-$7   & \hskip 7.0pt$-$9   & \hskip 
7.0pt$-$4	&   1  \\[2.8pt]
RXJ1419.3-2322    &  9.1 &  [T] & ATU	 & 1989.279 & 14 19 21.118  &$-$23 22 
13.20 & \hskip 3.5pt$+$40  & \hskip 3.5pt$-$30  & \hskip 3.5pt$+$24  & \hskip 
3.5pt$-$44  &  1,3 \\[2.8pt]
RXJ1447.3-3503    & 12.4 & [17] & AU	 & 1947.636 & 14 47 23.547  &$-$35 03 
11.51 & \hskip 3.5pt$-$25  & \hskip 3.5pt$-$45  & \hskip 3.5pt$-$43  & \hskip 
3.5pt$-$29  &      \\[2.8pt]
RXJ1448.2-4102    & 11.8 & [17] & AU	 & 1940.858 & 14 48 13.324  &$-$41 02 
57.77 & \hskip 3.5pt$-$24  & \hskip 3.5pt$-$31  & \hskip 3.5pt$-$35  & \hskip 
3.5pt$-$17  &      \\[2.8pt]
RXJ1450.4-3507    & 10.8 &  [T] & ATU	 & 1989.117 & 14 50 25.829  &$-$35 06 
48.47 & \hskip 3.5pt$-$15  & \hskip 3.5pt$-$18  & \hskip 3.5pt$-$21  & \hskip 
7.0pt$-$9	&   1  \\[2.8pt]
RXJ1450.5-3459    & 12.9 & [17] & AU	 & 1946.896 & 14 50 35.144  &$-$34 59 
04.28 & \hskip 3.5pt$-$17  & \hskip 3.5pt$-$32  & \hskip 3.5pt$-$31  & \hskip 
3.5pt$-$20  &      \\[2.8pt]
RXJ1452.4-3740    & 12.2 & [17] & AU	 & 1946.701 & 14 52 26.291  &$-$37 40 
07.82 & \hskip 3.5pt$-$11  & \hskip 3.5pt$-$18  & \hskip 3.5pt$-$18  & \hskip 
3.5pt$-$10  &      \\[2.8pt]
RXJ1454.2-3955    & 12.3 & [17] & AU	 & 1944.564 & 14 54 11.337  &$-$39 55 
22.12 & \hskip 3.5pt$-$22  & \hskip 3.5pt$-$30  & \hskip 3.5pt$-$33  & \hskip 
3.5pt$-$16  &      \\[2.8pt]
RXJ1457.3-3613    & 10.3 &  [T] & ATU	 & 1989.130 & 14 57 19.649  &$-$36 12 
27.15 & \hskip 3.5pt$-$27  & \hskip 3.5pt$-$20  & \hskip 3.5pt$-$34  & \hskip 
7.0pt$-$4	&  1,3 \\[2.8pt]
RXJ1458.6-3541    & 11.0 &  [T] & ATU	 & 1989.126 & 14 58 37.717  &$-$35 40 
30.19 & \hskip 3.5pt$-$21  & \hskip 3.5pt$-$25  & \hskip 3.5pt$-$31  & \hskip 
3.5pt$-$12  &   1  \\[2.8pt]
RXJ1458.7-3315    & 11.8 & [17] & AU	 & 1948.435 & 14 58 45.857  &$-$33 15 
09.44 & \hskip 7.0pt$-$6   & \hskip 3.5pt$-$27  & \hskip 3.5pt$-$18  & \hskip 
3.5pt$-$20  &      \\[2.8pt]
RXJ1459.3-4013    &  9.8 &  [T] & ATU	 & 1988.972 & 14 59 22.789  &$-$40 13 
11.79 & \hskip 3.5pt$-$27  & \hskip 3.5pt$-$26  & \hskip 3.5pt$-$36  & \hskip 
7.0pt$-$10  &   1  \\[2.8pt]
RXJ1500.6-4309    & 12.1 & [17] & AU	 & 1941.413 & 15 00 37.760  &$-$43 08 
32.52 & \hskip 3.5pt$-$24  & \hskip 3.5pt$-$19  & \hskip 3.5pt$-$30  & \hskip 
7.0pt$-$5	&      \\[2.8pt]
RXJ1500.8-4331    & 11.0 &  [T] & ATU	 & 1988.834 & 15 00 51.912  &$-$43 31 
20.93 & \hskip 3.5pt$-$21  & \hskip 3.5pt$-$19  & \hskip 3.5pt$-$27  & \hskip 
7.0pt$-$7	&   1  \\[2.8pt]
RXJ1501.2-4121    & 10.2 &  [T] & ATU	 & 1988.866 & 15 01 11.568  &$-$41 20 
40.40 & \hskip 3.5pt$-$15  & \hskip 3.5pt$-$16  & \hskip 3.5pt$-$21  & \hskip 
7.0pt$-$7	&   1  \\[2.8pt]
RXJ1502.4-3405    & 13.5 & [17] & AEU	 & 1956.751 & 15 02 26.069  &$-$34 05 
12.30 & \hskip 3.5pt$-$24  & \hskip 3.5pt$-$28  & \hskip 3.5pt$-$35  & \hskip 
3.5pt$-$12  &      \\[2.8pt]
RXJ1504.8-3950    &  9.7 &  [T] & AEHTU  & 1991.248 & 15 04 48.938  &$-$39 49 
23.36 & \hskip 3.5pt$-$35  & \hskip 3.5pt$-$36  & \hskip 3.5pt$-$49  & \hskip 
3.5pt$-$13  &  1,2 \\[2.8pt]
RXJ1505.4-3857    & 12.6 & [17] & AU	 & 1946.724 & 15 05 26.012  &$-$38 57 
00.84 & \hskip 3.5pt$-$24  & \hskip 7.0pt$+$1   & \hskip 3.5pt$-$20  & \hskip 
3.5pt$+$13  &      \\[2.8pt]
RXJ1505.9-4311    & 12.6 &  [7] & AU	 & 1940.819 & 15 05 56.988  &$-$43 12 
01.56 & \hskip 3.5pt$-$27  & \hskip 3.5pt$-$25  & \hskip 3.5pt$-$36  & \hskip 
7.0pt$-$8	&      \\[2.8pt]
RXJ1507.4-4601    & 11.7 &  [7] & AU	 & 1942.515 & 15 07 27.660  &$-$46 01 
06.25 & \hskip 3.5pt$-$33  & \hskip 3.5pt$-$18  & \hskip 3.5pt$-$37  & \hskip 
7.0pt$+$1	&      \\[2.8pt]
RXJ1507.6-4603    & 11.6 &  [7] & AU	 & 1942.348 & 15 07 37.875  &$-$46 03 
14.10 & \hskip 3.5pt$-$18  & \hskip 3.5pt$-$20  & \hskip 3.5pt$-$26  & \hskip 
7.0pt$-$8	&      \\[2.8pt]
RXJ1508.0-3338    & 13.1 & [17] & AEU	 & 1956.740 & 15 08 05.179  &$-$33 37 
54.42 & \hskip 3.5pt$-$31  & \hskip 3.5pt$-$26  & \hskip 3.5pt$-$40  & \hskip 
7.0pt$-$5	&      \\[2.8pt]
RXJ1508.6-4423    & 10.9 &  [T] & ATU	 & 1988.871 & 15 08 37.763  &$-$44 23 
16.76 & \hskip 3.5pt$-$18  & \hskip 3.5pt$-$17  & \hskip 3.5pt$-$24  & \hskip 
7.0pt$-$5	&   1  \\[2.8pt]
HD 133938         & 10.5 &  [T] & ATU	 & 1988.858 & 15 08 38.525  &$-$44 00 
51.87 & \hskip 3.5pt$-$20  & \hskip 3.5pt$-$24  & \hskip 3.5pt$-$30  & \hskip 
3.5pt$-$10  &   1  \\[2.8pt]
RXJ1508.8-3715    & 12.8 &  [7] & AEU	 & 1956.563 & 15 08 53.835  &$-$37 15 
45.76 & \hskip 3.5pt$-$26  & \hskip 3.5pt$-$31  & \hskip 3.5pt$-$39  & \hskip 
3.5pt$-$13  &      \\[2.8pt]
RXJ1509.3-4420    & 10.2 &  [T] & ATU	 & 1988.858 & 15 09 17.709  &$-$44 20 
09.80 & \hskip   14pt  0   & \hskip 7.0pt$-$5   & \hskip 7.0pt$-$2   & \hskip 
7.0pt$-$5	&   1  \\[2.8pt]
RXJ1511.0-3251    & 11.9 &  [7] & AEU	 & 1957.244 & 15 11 04.590  &$-$32 51 
29.17 & \hskip 7.0pt$-$8   & \hskip 3.5pt$-$13  & \hskip 3.5pt$-$14  & \hskip 
7.0pt$-$6	&      \\[2.8pt]
\hline
\end{tabular}
}
\end{table*}
\addtocounter{table}{-1}%
\begin{table*}[ht]
\scriptsize
{
\caption{-- {\it continued}}
\begin{tabular}{lccccccccccc}
\hline\\[-2.8pt]
Name              &  Mag & Ref. & Sources & Epoch    & 
\multispan2{\hfill{$\alpha$\hbox{\hskip 15pt} [J2000]\hbox{\hskip 15 pt} 
$\delta$}\hfill} & 
$\mu_{\alpha}{\rm cos}\delta$ & $\mu_{\delta}$ & $\mu_{l}{\rm cos}b$ & $\mu_{b}$ 
& Other PM's\\[2.8pt] 
\omit             & \omit& \omit& \omit   &\omit     & 
\multispan2{\hfill{$\overline{{\rm{[h\hskip 6pt m\hskip 6pt s]}}\hbox{\hskip 
50pt}\hbox{\hskip 
-12pt}[^{{\rm o}}\;\;\;\arcmin\;\;\;\arcsec]}$}\hfill} & 
\multispan4{\hfill{$\overline{\hbox{\hskip 60pt}{\rm [mas/yr]}\hbox{\hskip 
60pt}}$}\hfill} & 
\omit\\[2.8pt] 
\hline\\[-7.8pt]
RXJ1512.8-4508A   & 10.5 &  [T] & ATU	 & 1988.772 & 15 12 50.198  &$-$45 08 
04.28 & \hskip 3.5pt$-$22  & \hskip 3.5pt$-$21  & \hskip 3.5pt$-$29  & \hskip 
7.0pt$-$6	&      \\[2.8pt]
HD 135127         &  9.2 &  [T] & AETU   & 1988.783 & 15 14 39.598  &$-$34 45 
41.07 & \hskip 3.5pt$-$17  & \hskip 3.5pt$-$21  & \hskip 3.5pt$-$26  & \hskip 
7.0pt$-$8	&  1,3 \\[2.8pt]
RXJ1514.7-4220    & 11.3 & [17] & AU	 & 1941.111 & 15 14 47.597  &$-$42 20 
13.93 & \hskip 3.5pt$-$16  & \hskip 3.5pt$-$23  & \hskip 3.5pt$-$26  & \hskip 
7.0pt$-$11  &      \\[2.8pt]
RXJ1515.7-3332    & 11.6 & [17] & AEU	 & 1957.184 & 15 15 45.445  &$-$33 31 
58.63 & \hskip 3.5pt$-$25  & \hskip 3.5pt$-$32  & \hskip 3.5pt$-$39  & \hskip 
3.5pt$-$12  &      \\[2.8pt]
RXJ1515.8-4418    & 12.9 & [17] & AU	 & 1940.351 & 15 15 52.834  &$-$44 18 
16.21 & \hskip 3.5pt$-$31  & \hskip 3.5pt$-$36  & \hskip 3.5pt$-$46  & \hskip 
3.5pt$-$14  &      \\[2.8pt]
RXJ1516.6-4406    & 11.9 &  [7] & AU	 & 1940.086 & 15 16 36.824  &$-$44 07 
19.01 & \hskip 7.0pt$+$3   & \hskip 3.5pt$-$28  & \hskip 3.5pt$-$13  & \hskip 
3.5pt$-$25  &      \\[2.8pt]
RXJ1518.0-4445    & 12.3 & [17] & AU	 & 1940.142 & 15 18 01.393  &$-$44 44 
25.19 & \hskip 3.5pt$-$28  & \hskip 3.5pt$-$30  & \hskip 3.5pt$-$40  & \hskip 
3.5pt$-$10  &      \\[2.8pt]
RXJ1518.8-4050    & 11.0 &  [T] & ATU	 & 1988.738 & 15 18 52.841  &$-$40 50 
52.62 & \hskip 3.5pt$-$16  & \hskip 3.5pt$-$22  & \hskip 3.5pt$-$26  & \hskip 
3.5pt$-$10  &   1  \\[2.8pt]
RXJ1519.2-4056    & 11.4 &  [7] & AU	 & 1942.664 & 15 19 16.127  &$-$40 56 
05.98 & \hskip 3.5pt$-$23  & \hskip 3.5pt$-$35  & \hskip 3.5pt$-$39  & \hskip 
3.5pt$-$17  &      \\[2.8pt]
RXJ1519.6-4760    & 10.3 &  [T] & ATU	 & 1988.860 & 15 19 37.054  &$-$47 59 
33.83 & \hskip 3.5pt$-$30  & \hskip 3.5pt$-$32  & \hskip 3.5pt$-$43  & \hskip 
7.0pt$-$1	&   1  \\[2.8pt]
RXJ1522.2-3959    & 12.0 &  [7] & AU	 & 1944.956 & 15 22 11.752  &$-$39 59 
49.55 & \hskip 3.5pt$-$14  & \hskip 3.5pt$-$15  & \hskip 3.5pt$-$20  & \hskip 
7.0pt$-$5	&      \\[2.8pt]
RXJ1523.4-4055    & 11.9 &  [7] & AEU	 & 1952.572 & 15 23 25.667  &$-$40 55 
45.45 & \hskip 3.5pt$-$24  & \hskip 3.5pt$-$31  & \hskip 3.5pt$-$38  & \hskip 
3.5pt$-$12  &      \\[2.8pt]
RXJ1524.0-3209    & 12.4 &  [7] & AEU	 & 1955.736 & 15 24 03.115  &$-$32 09 
49.65 & \hskip 3.5pt$-$25  & \hskip 3.5pt$-$26  & \hskip 3.5pt$-$35  & \hskip 
7.0pt$-$6	&      \\[2.8pt]
RXJ1525.0-3604    & 11.0 &  [T] & AETU   & 1988.769 & 15 25 03.611  &$-$36 04 
45.25 & \hskip 3.5pt$-$12  & \hskip 3.5pt$-$27  & \hskip 3.5pt$-$25  & \hskip 
3.5pt$-$15  &   1  \\[2.8pt]
RXJ1525.5-3613    & 11.6 &  [7] & AEU	 & 1956.149 & 15 25 33.205  &$-$36 13 
45.87 & \hskip 3.5pt$-$17  & \hskip 3.5pt$-$28  & \hskip 3.5pt$-$30  & \hskip 
3.5pt$-$13  &      \\[2.8pt]
RXJ1526.0-4501    & 11.1 &  [T] & ATU	 & 1988.762 & 15 25 59.670  &$-$45 01 
15.49 & \hskip 3.5pt$-$21  & \hskip 3.5pt$-$19  & \hskip 3.5pt$-$28  & \hskip 
7.0pt$-$4	&   1  \\[2.8pt]
HD 137727         &  9.4 &  [T] & AHTU   & 1991.249 & 15 28 44.018  &$-$31 17 
38.39 & \hskip 3.5pt$+$21  & \hskip 3.5pt$+$56  & \hskip 3.5pt$+$51  & \hskip 
3.5pt$+$32  & 1,2,3\\[2.8pt]
RXJ1529.6-3546    & 10.6 &  [T] & AETU   & 1988.743 & 15 29 38.599  &$-$35 46 
51.12 & \hskip 3.5pt$-$23  & \hskip 3.5pt$-$24  & \hskip 3.5pt$-$32  & \hskip 
7.0pt$-$6	&  1,3 \\[2.8pt]
RXJ1529.7-3628    & 12.9 &  [7] & AEU	 & 1955.996 & 15 29 47.362  &$-$36 28 
36.44 & \hskip 3.5pt$-$33  & \hskip 3.5pt$-$35  & \hskip 3.5pt$-$47  & \hskip 
7.0pt$-$9	&      \\[2.8pt]
RXJ1529.8-4523    & 12.9 & [17] & AU	 & 1940.383 & 15 29 48.976  &$-$45 22 
44.39 & \hskip 3.5pt$-$17  & \hskip 3.5pt$-$30  & \hskip 3.5pt$-$31  & \hskip 
3.5pt$-$15  &      \\[2.8pt]
HD 138009         &  9.2 &  [T] & AEHTU  & 1991.248 & 15 30 26.256  &$-$32 18 
12.64 & \hskip 3.5pt$-$34  & \hskip 3.5pt$-$36  & \hskip 3.5pt$-$48  & \hskip 
7.0pt$-$8	& 1,2,3\\[2.8pt]
RXJ1531.3-3329    & 10.8 &  [T] & AETU   & 1988.736 & 15 31 21.950  &$-$33 29 
39.13 & \hskip 3.5pt$-$22  & \hskip 3.5pt$-$30  & \hskip 3.5pt$-$36  & \hskip 
3.5pt$-$10  &   1  \\[2.8pt]
RXJ1534.6-4003    & 11.9 & [17] & AEU	 & 1955.721 & 15 34 38.242  &$-$40 02 
26.98 & \hskip 3.5pt$-$36  & \hskip 3.5pt$-$40  & \hskip 3.5pt$-$53  & \hskip 
3.5pt$-$10  &      \\[2.8pt]
RXJ1538.0-3807    & 12.4 &  [7] & AEU	 & 1956.151 & 15 38 02.724  &$-$38 07 
22.17 & \hskip 3.5pt$-$22  & \hskip 3.5pt$-$22  & \hskip 3.5pt$-$31  & \hskip 
7.0pt$-$4	&      \\[2.8pt]
RXJ1538.6-3916    & 11.5 &  [7] & AEU	 & 1956.530 & 15 38 38.361  &$-$39 16 
54.08 & \hskip 3.5pt$-$19  & \hskip 3.5pt$-$40  & \hskip 3.5pt$-$40  & \hskip 
3.5pt$-$20  &      \\[2.8pt]
RXJ1538.7-4411    & 10.6 &  [T] & ATU	 & 1988.761 & 15 38 43.097  &$-$44 11 
47.06 & \hskip 3.5pt$-$24  & \hskip 3.5pt$-$29  & \hskip 3.5pt$-$36  & \hskip 
7.0pt$-$9	&   1  \\[2.8pt]
RXJ1539.7-3450    & 12.9 &  [7] & AEU	 & 1955.303 & 15 39 46.452  &$-$34 51 
01.64 & \hskip 3.5pt$-$26  & \hskip 3.5pt$-$20  & \hskip 3.5pt$-$33  & \hskip 
14pt    0	&      \\[2.8pt]
RXJ1540.0-4634    & 11.8 &  [7] & AU	 & 1942.991 & 15 40 02.592  &$-$46 34 
17.59 & \hskip 3.5pt$-$29  & \hskip 3.5pt$-$24  & \hskip 3.5pt$-$38  & \hskip 
7.0pt$-$2	&      \\[2.8pt]
RXJ1544.0-3311    & 10.8 &  [T] & AETU   & 1988.747 & 15 44 03.787  &$-$33 11 
10.88 & \hskip 3.5pt$-$21  & \hskip 3.5pt$-$27  & \hskip 3.5pt$-$34  & \hskip 
7.0pt$-$7	&   1  \\[2.8pt]
RXJ1544.8-3811    & 12.7 &  [7] & AEU	 & 1956.151 & 15 44 47.229  &$-$38 11 
39.47 & \hskip 3.5pt$-$16  & \hskip 3.5pt$-$25  & \hskip 3.5pt$-$28  & \hskip 
7.0pt$-$9	&      \\[2.8pt]
HT Lup            & 10.5 &  [T] & AEHT   & 1991.249 & 15 45 12.878  &$-$34 17 
30.44 & \hskip 7.0pt$-$3   & \hskip 3.5pt$-$58  & \hskip 3.5pt$-$39  & \hskip 
3.5pt$-$43  &  1,2 \\[2.8pt]
HD 140637         &  9.5 &  [T] & AHTU   & 1991.249 & 15 45 47.647  &$-$30 20 
54.87 & \hskip 3.5pt$-$70  &	       $-$103 & 	    $-$121 & \hskip 
3.5pt$-$34  & 1,2,3\\[2.8pt]
RXJ1545.9-4222    & 10.7 &  [T] & AETU   & 1988.573 & 15 45 52.260  &$-$42 22 
16.13 & \hskip 3.5pt$-$19  & \hskip 3.5pt$-$29  & \hskip 3.5pt$-$33  & \hskip 
3.5pt$-$11  &   1  \\[2.8pt]
RXJ1546.6-3618    & 11.3 &  [7] & AEU	 & 1956.074 & 15 46 41.251  &$-$36 18 
46.47 & \hskip 3.5pt$-$23  & \hskip 3.5pt$-$33  & \hskip 3.5pt$-$38  & \hskip 
3.5pt$-$11  &      \\[2.8pt]
RXJ1547.6-4018    & 11.1 &  [7] & AEU	 & 1954.644 & 15 47 41.848  &$-$40 18 
25.39 & \hskip 3.5pt$-$16  & \hskip 3.5pt$-$23  & \hskip 3.5pt$-$27  & \hskip 
7.0pt$-$8	&      \\[2.8pt]
RXJ1548.6-4335    & 11.2 &  [T] & ATU	 & 1988.860 & 15 48 42.029  &$-$43 35 
20.82 & \hskip 3.5pt$-$25  & \hskip 3.5pt$-$18  & \hskip 3.5pt$-$30  & \hskip 
7.0pt$+$1	&      \\[2.8pt]
GQ Lup            & 11.7 & [14] & AEU	 & 1955.986 & 15 49 12.144  &$-$35 39 
03.95 & \hskip 3.5pt$-$27  & \hskip 3.5pt$-$14  & \hskip 3.5pt$-$30  & \hskip 
7.0pt$+$6	&      \\[2.8pt]
HD 141277         & 10.7 &  [T] & AEHTU  & 1991.248 & 15 49 44.993  &$-$39 25 
08.89 & \hskip 3.5pt$-$20  & \hskip 3.5pt$-$25  & \hskip 3.5pt$-$31  & \hskip 
7.0pt$-$7	&  1,2 \\[2.8pt]
RXJ1549.9-3629    & 11.5 &  [7] & AEU	 & 1955.958 & 15 49 59.219  &$-$36 29 
56.59 & \hskip 3.5pt$-$30  & \hskip 3.5pt$-$21  & \hskip 3.5pt$-$36  & \hskip 
7.0pt$+$4	&      \\[2.8pt]
RXJ1550.1-4746    & 11.6 & [17] & AU	 & 1942.143 & 15 50 12.047  &$-$47 46 
09.95 & \hskip 3.5pt$-$13  & \hskip 7.0pt$-$3   & \hskip 3.5pt$-$12  & \hskip 
7.0pt$+$5	&      \\[2.8pt]
Sz 77             & 12.5 & [13] & AEU	 & 1955.966 & 15 51 46.992  &$-$35 56 
42.84 & \hskip 7.0pt$-$7   & \hskip 3.5pt$-$30  & \hskip 3.5pt$-$25  & \hskip 
3.5pt$-$19  &      \\[2.8pt]
RXJ1555.4-3338    & 12.4 &  [7] & AU	 & 1947.721 & 15 55 26.311  &$-$33 38 
21.67 & \hskip 3.5pt$-$24  & \hskip 3.5pt$-$34  & \hskip 3.5pt$-$41  & \hskip 
3.5pt$-$10  &      \\[2.8pt]
RU Lup            & 11.5 &  [T] & AEHTU  & 1991.248 & 15 56 42.320  &$-$37 49 
15.27 & \hskip 3.5pt$-$11  & \hskip 3.5pt$-$26  & \hskip 3.5pt$-$25  & \hskip 
3.5pt$-$13  &  1,2 \\[2.8pt]
Sz 129            & 13.0 & [13] & AU	 & 1939.940 & 15 59 16.541  &$-$41 57 
09.18 & \hskip 3.5pt$-$14  & \hskip 3.5pt$-$21  & \hskip 3.5pt$-$25  & \hskip 
7.0pt$-$7	&      \\[2.8pt]
RY Lup            & 11.2 &  [T] & AEHTU  & 1991.248 & 15 59 28.395  &$-$40 21 
51.05 & \hskip 3.5pt$-$13  & \hskip 3.5pt$-$20  & \hskip 3.5pt$-$23  & \hskip 
7.0pt$-$7	&  1,2 \\[2.8pt]
CD-36 10569       & 11.0 &  [T] & AHTU   & 1991.249 & 15 59 49.531  &$-$36 28 
27.53 & \hskip 3.5pt$-$29  & \hskip 3.5pt$-$43  & \hskip 3.5pt$-$50  & \hskip 
3.5pt$-$13  &  1,2 \\[2.8pt]
CD-41 10484       & 11.6 & [14] & AU	 & 1940.164 & 16 00 44.600  &$-$41 55 
29.55 & \hskip 3.5pt$-$21  & \hskip 3.5pt$-$24  & \hskip 3.5pt$-$32  & \hskip 
7.0pt$-$4	&      \\[2.8pt]
RXJ1601.1-3320    & 10.9 &  [T] & ATU	 & 1989.061 & 16 01 08.981  &$-$33 20 
14.02 & \hskip 3.5pt$-$11  & \hskip 3.5pt$-$20  & \hskip 3.5pt$-$22  & \hskip 
7.0pt$-$7	&   1  \\[2.8pt]
RXJ1601.8-4026    & 13.5 &  [7] & AEU	 & 1951.768 & 16 01 49.497  &$-$40 26 
18.41 & \hskip 3.5pt$-$15  & \hskip 3.5pt$-$13  & \hskip 3.5pt$-$20  & \hskip 
14pt    0	&      \\[2.8pt]
RXJ1601.9-3613    & 12.0 &  [7] & AU	 & 1946.725 & 16 01 59.272  &$-$36 12 
53.89 & \hskip 3.5pt$-$21  & \hskip 3.5pt$-$36  & \hskip 3.5pt$-$40  & \hskip 
3.5pt$-$13  &      \\[2.8pt]
HD 143677         &  9.8 &  [T] & AHTU   & 1991.248 & 16 03 45.379  &$-$43 55 
49.04 & \hskip 3.5pt$-$14  & \hskip 3.5pt$-$22  & \hskip 3.5pt$-$25  & \hskip 
7.0pt$-$8	& 1,2,3\\[2.8pt]
RXJ1603.8-3938    & 11.2 &  [T] & AETU   & 1988.576 & 16 03 52.499  &$-$39 39 
00.91 & \hskip 3.5pt$-$24  & \hskip 3.5pt$-$28  & \hskip 3.5pt$-$36  & \hskip 
7.0pt$-$4	&   1  \\[2.8pt]
RXJ1605.7-3905    & 10.7 &  [T] & AETUV  & 1992.899 & 16 05 45.012  &$-$39 06 
06.34 & \hskip 3.5pt$-$18  & \hskip 3.5pt$-$29  & \hskip 3.5pt$-$33  & \hskip 
7.0pt$-$9	&   1  \\[2.8pt]
RXJ1606.3-4447    & 12.3 &  [7] & AU	 & 1941.982 & 16 06 23.446  &$-$44 47 
34.09 & \hskip 3.5pt$-$24  & \hskip 3.5pt$-$31  & \hskip 3.5pt$-$39  & \hskip 
7.0pt$-$7	&      \\[2.8pt]
HO Lup            & 13.4 &  [V] & AEUV   & 1995.334 & 16 07 00.605  &$-$39 02 
19.37 & \hskip 3.5pt$-$14  & \hskip 3.5pt$-$29  & \hskip 3.5pt$-$30  & \hskip 
3.5pt$-$12  &      \\[2.8pt]
Sz 90             & 15.0 &  [V] & EUV	 & 1998.400 & 16 07 10.077  &$-$39 11 
03.18 & \hskip 7.0pt$+$5   & \hskip 3.5pt$-$24  & \hskip 3.5pt$-$12  & \hskip 
3.5pt$-$21  &      \\[2.8pt]
Sz 91             & 15.9 &  [V] & EUV	 & 1998.400 & 16 07 11.613  &$-$39 03 
47.08 & \hskip 7.0pt$-$1   & \hskip 7.0pt$+$4   & \hskip 7.0pt$+$2   & \hskip 
7.0pt$+$4	&      \\[2.8pt]
RXJ1608.0-3857    & 13.2 &  [V] & EUV	 & 1998.400 & 16 07 59.963  &$-$38 57 
50.91 & \hskip 7.0pt$-$8   & \hskip 3.5pt$-$24  & \hskip 3.5pt$-$22  & \hskip 
3.5pt$-$12  &      \\[2.8pt]
\hline
\end{tabular}
}
\end{table*}
\addtocounter{table}{-1}%
\begin{table*}[ht]
\scriptsize
{
\caption{ -- {\it continued}}
\begin{tabular}{lccccccccccc}
\hline\\[-2.8pt]
Name              &  Mag & Ref. & Sources & Epoch    & 
\multispan2{\hfill{$\alpha$\hbox{\hskip 15pt} [J2000]\hbox{\hskip 15 pt} 
$\delta$}\hfill} & 
$\mu_{\alpha}{\rm cos}\delta$ & $\mu_{\delta}$ & $\mu_{l}{\rm cos}b$ & $\mu_{b}$ 
& Other PM's\\[2.8pt] 
\omit             & \omit& \omit& \omit   &\omit     & 
\multispan2{\hfill{$\overline{{\rm{[h\hskip 6pt m\hskip 6pt s]}}\hbox{\hskip 
50pt}\hbox{\hskip 
-12pt}[^{{\rm o}}\;\;\;\arcmin\;\;\;\arcsec]}$}\hfill} & 
\multispan4{\hfill{$\overline{\hbox{\hskip 60pt}{\rm [mas/yr]}\hbox{\hskip 
60pt}}$}\hfill} & 
\omit\\[2.8pt] 
\hline\\[-7.8pt]
F 304             & 13.1 &  [V] & EUV	 & 1998.400 & 16 08 10.959  &$-$39 10 
45.85 & \hskip 3.5pt$-$13  & \hskip 3.5pt$-$18  & \hskip 3.5pt$-$22  & \hskip 
7.0pt$-$4	&      \\[2.8pt]
Sz 96             & 14.2 &  [V] & EUV	 & 1998.400 & 16 08 12.636  &$-$39 08 
33.26 & \hskip 7.0pt$-$3   & \hskip 3.5pt$-$11  & \hskip 3.5pt$-$10  & \hskip 
7.0pt$-$6	&      \\[2.8pt]
RXJ1608.3-3843    & 12.2 &  [7] & AE	 & 1943.984 & 16 08 18.322  &$-$38 44 
03.61 & \hskip 3.5pt$-$16  & \hskip 3.5pt$-$38  & \hskip 3.5pt$-$38  & \hskip 
3.5pt$-$17  &      \\[2.8pt]
HK Lup            & 14.7 &  [V] & UV	 & 1999.284 & 16 08 22.499  &$-$39 04 
46.32 & \hskip 3.5pt$-$13  & \hskip 3.5pt$-$10  & \hskip 3.5pt$-$16  & \hskip 
7.0pt$+$2	&      \\[2.8pt]
RXJ1608.4-3900A,B & 15.1 &  [V] & EUV	 & 1998.400 & 16 08 27.791  &$-$39 00 
40.62 & \hskip 7.0pt$+$1   & \hskip 3.5pt$-$18  & \hskip 3.5pt$-$12  & \hskip 
3.5pt$-$14  &      \\[2.8pt]
Sz 108            & 13.4 &  [V] & DV	 & 1997.977 & 16 08 42.739  &$-$39 06 
18.32 & \hskip 7.0pt$-$8   & \hskip 3.5pt$-$51  & \hskip 3.5pt$-$41  & \hskip 
3.5pt$-$32  &      \\[2.8pt]
Sz 110            & 15.2 &  [V] & EUV	 & 1998.400 & 16 08 51.561  &$-$39 03 
17.53 & \hskip 7.0pt$+$3   & \hskip 3.5pt$-$20  & \hskip 3.5pt$-$12  & \hskip 
3.5pt$-$17  &      \\[2.8pt]
RXJ1608.9-3905    & 12.2 &  [T] & ATUV   & 1993.163 & 16 08 54.276  &$-$39 06 
05.66 & \hskip 3.5pt$-$10  & \hskip 3.5pt$-$26  & \hskip 3.5pt$-$25  & \hskip 
3.5pt$-$12  &   1  \\[2.8pt]
V908 Sco          & 14.7 &  [V] & EUV	 & 1998.400 & 16 09 01.838  &$-$39 05 
12.13 & \hskip 3.5pt$-$10  & \hskip 3.5pt$-$16  & \hskip 3.5pt$-$19  & \hskip 
7.0pt$-$5	&      \\[2.8pt]
RXJ1609.3-3855    & 15.5 &  [V] & EUV	 & 1998.400 & 16 09 23.194  &$-$38 55 
54.80 & \hskip 3.5pt$+$30  & \hskip 3.5pt$-$68  & \hskip 3.5pt$-$25  & \hskip 
3.5pt$-$70  &      \\[2.8pt]
RXJ1609.6-3854    & 11.3 &  [V] & EUV	 & 1998.400 & 16 09 39.526  &$-$38 55 
06.98 & \hskip 3.5pt$-$12  & \hskip 3.5pt$-$16  & \hskip 3.5pt$-$20  & \hskip 
7.0pt$-$4	&      \\[2.8pt]
Sz 119            & 14.8 &  [V] & EUV	 & 1998.400 & 16 09 57.059  &$-$38 59 
47.65 & \hskip 3.5pt$-$38  & \hskip 3.5pt$-$59  & \hskip 3.5pt$-$68  & \hskip 
3.5pt$-$17  &      \\[2.8pt]
RXJ1610.0-4016    & 11.2 &  [7] & AEU	 & 1952.866 & 16 10 04.869  &$-$40 16 
11.06 & \hskip 3.5pt$-$20  & \hskip 3.5pt$-$32  & \hskip 3.5pt$-$37  & \hskip 
7.0pt$-$9	&      \\[2.8pt]
Sz 122            & 14.5 &  [V] & EUV	 & 1998.400 & 16 10 16.435  &$-$39 08 
00.50 & \hskip 3.5pt$-$10  & \hskip 3.5pt$-$30  & \hskip 3.5pt$-$28  & \hskip 
3.5pt$-$15  &      \\[2.8pt]
RXJ1611.2-3905    & 15.6 &  [V] & EUV	 & 1998.400 & 16 11 12.898  &$-$39 05 
13.15 & \hskip 3.5pt$-$25  & \hskip 7.0pt$-$6   & \hskip 3.5pt$-$23  & \hskip 
3.5pt$+$13  &      \\[2.8pt]
Sz 124            & 13.5 &  [V] & EUV	 & 1998.400 & 16 11 53.352  &$-$39 02 
15.69 & \hskip 3.5pt$-$16  & \hskip 3.5pt$-$20  & \hskip 3.5pt$-$25  & \hskip 
7.0pt$-$4	&      \\[2.8pt]
RXJ1612.0-3840    & 11.7 &  [7] & AEU	 & 1955.181 & 16 12 01.429  &$-$38 40 
26.53 & \hskip 7.0pt$-$5   & \hskip 3.5pt$-$33  & \hskip 3.5pt$-$26  & \hskip 
3.5pt$-$20  &      \\[2.8pt]
RXJ1613.1-3804    & 12.9 &  [7] & AEU	 & 1956.651 & 16 13 12.757  &$-$38 03 
50.21 & \hskip 3.5pt$-$13  & \hskip 3.5pt$-$24  & \hskip 3.5pt$-$26  & \hskip 
7.0pt$-$8	&      \\[2.8pt]
RXJ1615.9-3241    & 13.0 &  [7] & AU	 & 1948.777 & 16 15 57.089  &$-$32 41 
23.10 & \hskip 3.5pt$-$17  & \hskip 3.5pt$-$41  & \hskip 3.5pt$-$41  & \hskip 
3.5pt$-$17  &      \\[2.8pt]
HD 147048         & 10.7 &  [T] & AETU   & 1988.512 & 16 21 12.205  &$-$40 30 
20.27 & \hskip 3.5pt$-$13  & \hskip 3.5pt$-$26  & \hskip 3.5pt$-$27  & \hskip 
7.0pt$-$9	&  1,3 \\[2.8pt]
HD 147402         & 10.7 &  [T] & ATU	 & 1988.889 & 16 23 29.559  &$-$39 58 
00.48 & \hskip 3.5pt$-$13  & \hskip 3.5pt$-$24  & \hskip 3.5pt$-$26  & \hskip 
7.0pt$-$8	&  1,3 \\[2.8pt]
HD 147454         &  9.3 &  [T] & ATU	 & 1989.084 & 16 23 32.322  &$-$34 39 
49.67 & \hskip 3.5pt$-$20  & \hskip 3.5pt$-$24  & \hskip 3.5pt$-$31  & \hskip 
7.0pt$-$2	&  1,3 \\[2.8pt]
SAO 207620        &  9.5 &  [T] & ATU	 & 1989.084 & 16 23 37.654  &$-$34 40 
21.50 & \hskip 3.5pt$-$16  & \hskip 3.5pt$-$31  & \hskip 3.5pt$-$33  & \hskip 
3.5pt$-$10  &  1,3 \\[2.8pt]
\hline
U--Sco - Oph\\[2.8pt]
\hline
GSC 6770-0655     & 12.6 & [10] & AU	 & 1946.915 & 15 19 53.057  &$-$28 02 
23.61 & \hskip 3.5pt$-$47  & \hskip 3.5pt$-$32  & \hskip 3.5pt$-$57  & \hskip 
7.0pt$+$2	&   5  \\[2.8pt]
GSC 6764-1305     & 13.1 & [10] & AU	 & 1951.068 & 15 35 57.866  &$-$23 24 
03.83 & \hskip 3.5pt$-$20  & \hskip 3.5pt$-$24  & \hskip 3.5pt$-$31  & \hskip 
7.0pt$-$6	&   5  \\[2.8pt]
GSC 6195-0768     & 12.7 & [10] & AU	 & 1949.666 & 15 57 02.402  &$-$19 50 
40.83 & \hskip 3.5pt$-$28  & \hskip 3.5pt$-$32  & \hskip 3.5pt$-$42  & \hskip 
7.0pt$-$3	&   5  \\[2.8pt]
He 3-1126         & 10.3 &  [T] & ATU	 & 1989.313 & 15 58 36.923  &$-$22 57 
15.02 & \hskip 3.5pt$-$13  & \hskip 3.5pt$-$19  & \hskip 3.5pt$-$21  & \hskip 
7.0pt$-$6	&  1,3 \\[2.8pt]
GSC 6191-0552     & 12.9 & [10] & AU	 & 1948.585 & 15 58 47.781  &$-$17 57 
58.74 & \hskip 3.5pt$-$20  & \hskip 3.5pt$-$18  & \hskip 3.5pt$-$26  & \hskip 
7.0pt$+$2	&   5  \\[2.8pt]
GSC 6191-0019     & 13.0 & [10] & AU	 & 1948.599 & 15 59 02.082  &$-$18 44 
12.75 & \hskip 7.0pt$-$2   & \hskip 3.5pt$-$34  & \hskip 3.5pt$-$26  & \hskip 
3.5pt$-$22  &   5  \\[2.8pt]
GSC 6784-1219     & 11.3 &  [T] & ATU	 & 1989.063 & 16 05 50.692  &$-$25 33 
13.58 & \hskip 3.5pt$-$22  & \hskip 3.5pt$-$36  & \hskip 3.5pt$-$41  & \hskip 
3.5pt$-$10  &   5  \\[2.8pt]
GSC 6780-1061     & 13.2 &  [V] & UV	 & 1999.263 & 16 06 54.381  &$-$24 16 
10.95 & \hskip 7.0pt$+$3   & \hskip 3.5pt$-$17  & \hskip 3.5pt$-$10  & \hskip 
3.5pt$-$14  &      \\[2.8pt]
GSC 6784-0039     & 10.3 &  [T] & ATU	 & 1989.068 & 16 08 43.419  &$-$26 02 
16.55 & \hskip 3.5pt$-$12  & \hskip 3.5pt$-$25  & \hskip 3.5pt$-$27  & \hskip 
7.0pt$-$9	&  1,5 \\[2.8pt]
GSC 6213-0194     & 13.8 & [10] & AU	 & 1950.614 & 16 09 41.040  &$-$22 17 
58.34 & \hskip 3.5pt$-$20  & \hskip 3.5pt$-$30  & \hskip 3.5pt$-$36  & \hskip 
7.0pt$-$6	&   5  \\[2.8pt]
Wa Oph/1          & 11.6 & [15] & AU	 & 1933.391 & 16 11 08.986  &$-$19 04 
45.33 & \hskip 3.5pt$-$35  & \hskip 3.5pt$-$20  & \hskip 3.5pt$-$38  & \hskip 
3.5pt$+$12  &      \\[2.8pt]
V866 Sco A,B      & 12.4 &  [2] & AU	 & 1948.893 & 16 11 31.402  &$-$18 38 
24.54 & \hskip 3.5pt$-$17  & \hskip 3.5pt$-$31  & \hskip 3.5pt$-$34  & \hskip 
7.0pt$-$8	&      \\[2.8pt]
Wa Oph/2          & 11.6 & [15] & AU	 & 1948.893 & 16 11 59.330  &$-$19 06 
52.40 & \hskip 3.5pt$-$26  & \hskip 3.5pt$-$19  & \hskip 3.5pt$-$31  & \hskip 
7.0pt$+$6	&      \\[2.8pt]
Wa Oph/3          & 11.0 &  [T] & ATU	 & 1988.446 & 16 12 40.514  &$-$18 59 
27.87 & \hskip 3.5pt$-$12  & \hskip 3.5pt$-$28  & \hskip 3.5pt$-$29  & \hskip 
3.5pt$-$10  &  1,3 \\[2.8pt]
GSC 6793-0797     & 12.6 & [10] & AU	 & 1950.604 & 16 13 02.770  &$-$22 57 
43.43 & \hskip 3.5pt$-$17  & \hskip 3.5pt$-$37  & \hskip 3.5pt$-$39  & \hskip 
3.5pt$-$13  &   5  \\[2.8pt]
GSC 6213-0306     & 10.6 &  [T] & ATU	 & 1989.287 & 16 13 18.593  &$-$22 12 
48.77 & \hskip 3.5pt$-$14  & \hskip 3.5pt$-$22  & \hskip 3.5pt$-$25  & \hskip 
7.0pt$-$4	&  1,5 \\[2.8pt]
GSC 6793-0569     & 12.5 & [10] & AU	 & 1951.086 & 16 13 29.333  &$-$23 11 
06.16 & \hskip 3.5pt$-$17  & \hskip 3.5pt$-$37  & \hskip 3.5pt$-$39  & \hskip 
3.5pt$-$13  &   5  \\[2.8pt]
GSC 6793-0994     & 12.2 & [10] & AU	 & 1951.086 & 16 14 02.142  &$-$23 01 
01.15 & \hskip 3.5pt$-$15  & \hskip 3.5pt$-$28  & \hskip 3.5pt$-$31  & \hskip 
7.0pt$-$8	&   5  \\[2.8pt]
PDS 81            & 11.8 & [14] & AU	 & 1934.306 & 16 14 07.975  &$-$19 38 
27.67 & \hskip 7.0pt$-$5   & \hskip 7.0pt$-$8   & \hskip 3.5pt$-$10  & \hskip 
7.0pt$-$1	&      \\[2.8pt]
GSC 6793-0819     & 10.9 &  [T] & ATU	 & 1989.301 & 16 14 11.080  &$-$23 05 
35.84 & \hskip 3.5pt$-$14  & \hskip 3.5pt$-$22  & \hskip 3.5pt$-$25  & \hskip 
7.0pt$-$5	&  1,5 \\[2.8pt]
GSC 6801-0186     & 11.8 & [10] & AU	 & 1946.102 & 16 14 59.251  &$-$27 50 
21.76 & \hskip 3.5pt$-$14  & \hskip 3.5pt$-$29  & \hskip 3.5pt$-$31  & \hskip 
3.5pt$-$11  &   5  \\[2.8pt]
GSC 6793-1406     & 10.8 &  [T] & ATU	 & 1989.204 & 16 16 17.951  &$-$23 39 
47.30 & \hskip 3.5pt$-$11  & \hskip 3.5pt$-$24  & \hskip 3.5pt$-$26  & \hskip 
7.0pt$-$8	& 1,3,5\\[2.8pt]
GSC 6214-2384     & 12.0 & [10] & AEU	 & 1960.860 & 16 19 33.997  &$-$22 28 
28.52 & \hskip 3.5pt$-$23  & \hskip 3.5pt$-$34  & \hskip 3.5pt$-$40  & \hskip 
7.0pt$-$6	&   5  \\[2.8pt]
RXJ1621.4-2332ab  & 11.6 & [A]  & AEU	 & 1960.318 & 16 21 28.819  &$-$23 32 
38.49 & \hskip 3.5pt$-$13  & \hskip 3.5pt$-$32  & \hskip 3.5pt$-$33  & \hskip 
3.5pt$-$12  &      \\[2.8pt]
GSC 6214-0210     & 13.1 & [10] & AU	 & 1938.443 & 16 21 54.727  &$-$20 43 
06.64 & \hskip 7.0pt$+$6   & \hskip 3.5pt$-$39  & \hskip 3.5pt$-$25  & \hskip 
3.5pt$-$30  &   5  \\[2.8pt]
GSC 6794-0537     & 12.7 & [10] & AEU	 & 1961.643 & 16 23 07.847  &$-$23 00 
59.03 & \hskip 3.5pt$-$11  & \hskip 3.5pt$-$28  & \hskip 3.5pt$-$28  & \hskip 
3.5pt$-$11  &   5  \\[2.8pt]
GSC 6798-0035     & 12.1 & [10] & AEU	 & 1957.463 & 16 23 32.364  &$-$25 23 
48.01 & \hskip 3.5pt$-$15  & \hskip 3.5pt$-$19  & \hskip 3.5pt$-$24  & \hskip 
7.0pt$-$1	&   5  \\[2.8pt]
RXJ1624.8-2359    & 14.2 &  [V] & EUV	 & 1998.539 & 16 24 48.446  &$-$23 59 
16.58 & \hskip 3.5pt$+$32  & \hskip 3.5pt$-$59  & \hskip 3.5pt$-$23  & \hskip 
3.5pt$-$63  &      \\[2.8pt]
GSC 6794-0156     &  9.9 &  [T] & AETU   & 1989.138 & 16 24 51.377  &$-$22 39 
32.32 & \hskip 3.5pt$-$14  & \hskip 3.5pt$-$25  & \hskip 3.5pt$-$29  & \hskip 
7.0pt$-$6	& 1,3,5\\[2.8pt]
Haro 1-4          & 13.4 & [11] & AEU	 & 1962.905 & 16 25 10.544  &$-$23 19 
13.43 & \hskip 3.5pt$-$15  & \hskip 3.5pt$-$29  & \hskip 3.5pt$-$32  & \hskip 
7.0pt$-$8	&      \\[2.8pt]
RXJ1625.3-2402    & 15.8 &  [V] & EUV	 & 1998.539 & 16 25 22.484  &$-$24 02 
06.35 & \hskip 3.5pt$+$31  & \hskip 3.5pt$-$53  & \hskip 3.5pt$-$19  & \hskip 
3.5pt$-$58  &      \\[2.8pt]
RXJ1625.4-2346    & 11.6 &  [A] & AEU	 & 1962.905 & 16 25 28.637  &$-$23 46 
25.64 & \hskip 3.5pt$-$14  & \hskip 3.5pt$-$19  & \hskip 3.5pt$-$23  & \hskip 
7.0pt$-$2	&      \\[2.8pt]
DoAr 22           & 13.0 &  [5] & AEU	 & 1960.240 & 16 26 19.331  &$-$23 43 
19.70 & \hskip   14pt  0   & \hskip 3.5pt$-$29  & \hskip 3.5pt$-$22  & \hskip 
3.5pt$-$19  &      \\[2.8pt]
GSC 6794-0337     & 12.3 & [10] & AEU	 & 1961.422 & 16 27 39.603  &$-$22 45 
21.85 & \hskip 3.5pt$-$15  & \hskip 3.5pt$-$25  & \hskip 3.5pt$-$29  & \hskip 
7.0pt$-$4	&   5  \\[2.8pt]
SR 9              & 11.3 & [11] & AEU	 & 1957.330 & 16 27 40.324  &$-$24 22 
02.68 & \hskip 3.5pt$-$15  & \hskip 3.5pt$-$40  & \hskip 3.5pt$-$40  & \hskip 
3.5pt$-$15  &      \\[2.8pt]
Haro 1-14/c       & 12.2 &  [V] & AEUV   & 1995.429 & 16 31 04.397  &$-$24 04 
33.38 & \hskip 7.0pt$-$4   & \hskip 3.5pt$-$39  & \hskip 3.5pt$-$33  & \hskip 
3.5pt$-$22  &      \\[2.8pt]
Haro 1-14         & 14.6 &  [V] & EUV    & 1998.539 & 16 31 05.199  &$-$24 04 
40.44 &		  \hskip 3.5pt$+$38   & \hskip 3.5pt$-$59  & \hskip 3.5pt$-$20  & \hskip 
3.5pt$-$67  &      \\[2.8pt]
CD-24 12809       & 11.0 &  [T] & AETU   & 1988.787 & 16 45 26.158  &$-$25 03 
16.44 & \hskip 7.0pt$-$1   & \hskip 3.5pt$-$23  & \hskip 3.5pt$-$18  & \hskip 
3.5pt$-$14  &   1  \\[2.8pt]
\hline
\end{tabular}
}
\end{table*}
\addtocounter{table}{-1}%
\begin{table*}[ttt]
\scriptsize
{
\caption{ -- {\it continued}}
\begin{tabular}{lccccccccccc}
\hline\\[-2.8pt]
Name              &  Mag & Ref. & Sources & Epoch    & 
\multispan2{\hfill{$\alpha$\hbox{\hskip 15pt} [J2000]\hbox{\hskip 15 pt} 
$\delta$}\hfill} & 
$\mu_{\alpha}{\rm cos}\delta$ & $\mu_{\delta}$ & $\mu_{l}{\rm cos}b$ & $\mu_{b}$ 
& Other PM's\\[2.8pt] 
\omit             & \omit& \omit& \omit   &\omit     & 
\multispan2{\hfill{$\overline{{\rm{[h\hskip 6pt m\hskip 6pt s]}}\hbox{\hskip 
50pt}\hbox{\hskip 
-12pt}[^{{\rm o}}\;\;\;\arcmin\;\;\;\arcsec]}$}\hfill} & 
\multispan4{\hfill{$\overline{\hbox{\hskip 60pt}{\rm [mas/yr]}\hbox{\hskip 
60pt}}$}\hfill} & 
\omit\\[2.8pt] 
\hline\\[-7.8pt]
PDS 89            & 12.0 & [14] & AU	 & 1930.004 & 16 47 13.617  &$-$15 14 
25.94 & \hskip 7.0pt$-$5   & \hskip 3.5pt$-$11  & \hskip 3.5pt$-$12  & \hskip 
7.0pt$-$2	&      \\[2.8pt]
V1121 Oph         & 11.4 &  [H] & AHU	 & 1991.248 & 16 49 15.308  &$-$14 22 
08.43 & \hskip 3.5pt$-$10  & \hskip 3.5pt$-$21  & \hskip 3.5pt$-$23  & \hskip 
7.0pt$-$4	&  1,2 \\[2.8pt]
AS 216            & 11.0 &  [H] & AHU	 & 1991.249 & 17 00 34.980  &$-$27 38 
04.23 & \hskip 7.0pt$+$3   & \hskip 3.5pt$-$23  & \hskip 3.5pt$-$17  & \hskip 
3.5pt$-$16  &  1,2 \\[2.8pt]
IX Oph            & 11.0 &  [H] & AEH	 & 1991.249 & 17 09 48.177  &$-$27 16 
59.21 & \hskip 7.0pt$-$21  & \hskip 7.0pt$-$7   & \hskip 3.5pt$-$18  & \hskip 
3.5pt$+$13  &  1,2 \\[2.8pt]
\hline
Corona Australis\\[2.8pt]
\hline
CrAPMS 4SE        & 10.9 & [16] & AEU	 & 1956.051 & 18 57 20.814  &$-$36 42 
59.45 & \hskip 7.0pt$+$4   & \hskip 3.5pt$-$34  & \hskip 3.5pt$-$30  & \hskip 
3.5pt$-$16  &      \\[2.8pt]
CrAPMS 5          & 12.3 & [16] & AEU	 & 1956.055 & 18 58 01.834  &$-$36 53 
43.83 & \hskip 3.5pt$+$13  & \hskip 3.5pt$-$17  & \hskip 3.5pt$-$11  & \hskip 
3.5pt$-$18  &      \\[2.8pt]
Wa CrA/2          & 10.6 &  [T] & AETU   & 1988.788 & 19 02 01.981  &$-$37 07 
43.19 & \hskip 7.0pt$+$3   & \hskip 3.5pt$-$29  & \hskip 3.5pt$-$26  & \hskip 
3.5pt$-$13  &   1  \\[2.8pt]
\hline
Other  regions\\[2.8pt]
\hline
V4046 Sgr         & 10.6 &  [T] & ATU	 & 1989.126 & 18 14 10.472  &$-$32 47 
33.89 & \hskip 7.0pt$+$1   & \hskip 3.5pt$-$54  & \hskip 3.5pt$-$48  & \hskip 
3.5pt$-$26  &   1  \\[2.8pt]
FK Ser            & 10.7 &  [H] & AH	 & 1991.249 & 18 20 22.744  &$-$10 11 
13.33 & \hskip 7.0pt$-$1   & \hskip 3.5pt$-$39  & \hskip 3.5pt$-$35  & \hskip 
3.5pt$-$18  &   2  \\[2.8pt]
\hline
\end{tabular}
}
See references in Table \ref{Tab6}.
\end{table*}

\begin{table*}[ht]
\scriptsize
{
\caption{\label{Tab5}Proper motions for the HAeBe stars}
\begin{tabular}{lccccccccccc}
\hline\\[-2.8pt]
Name              &  Mag & Ref. & Sources & Epoch    & 
\multispan2{\hfill{$\alpha$\hbox{\hskip 15pt} [J2000]\hbox{\hskip 15 pt} 
$\delta$}\hfill} & 
$\mu_{\alpha}{\rm cos}\delta$ & $\mu_{\delta}$ & $\mu_{l}{\rm cos}b$ & $\mu_{b}$ 
& Other PM's\\[2.8pt] 
\omit             & \omit& \omit& \omit   &\omit     & 
\multispan2{\hfill{$\overline{{\rm{[h\hskip 6pt m\hskip 6pt s]}}\hbox{\hskip 
50pt}\hbox{\hskip 
-12pt}[^{{\rm o}}\;\;\;\arcmin\;\;\;\arcsec]}$}\hfill} & 
\multispan4{\hfill{$\overline{\hbox{\hskip 45pt}{\rm [mas/yr]}\hbox{\hskip 
45pt}}$}\hfill} & 
\omit\\[2.8pt] 
\hline\\[-7.8pt]
Chamaeleon\\[2.8pt]
\hline
HD 96675          &  7.7 &  [T] & AEHT   & 1991.248 & 11 05 57.865  &$-$76 07 
48.89 & \hskip 3.5pt$-$24  & \hskip  14pt   0   & \hskip 3.5pt$-$22  & \hskip 
3.5pt$-$10  & 1,2,3\\[2.8pt]
CU Cha            &  8.5 &  [T] & AHT	 & 1991.249 & 11 08 03.372  &$-$77 39 
17.50 & \hskip 3.5pt$-$22  & \hskip 7.0pt$+$1   & \hskip 3.5pt$-$20  & \hskip 
7.0pt$-$8	& 1,2,3\\[2.8pt]
HD 97300          &  9.0 &  [T] & AHTUV  & 1991.252 & 11 09 50.073  &$-$76 36 
47.71 & \hskip 3.5pt$-$22  & \hskip 3.5pt$+$17  & \hskip 3.5pt$-$27  & \hskip 
7.0pt$+$7	& 1,2,3\\[2.8pt]
HD 102065         &  6.6 &  [T] & AHTU   & 1991.248 & 11 43 37.057  &$-$80 29 
00.47 & \hskip 3.5pt$-$28  & \hskip 7.0pt$-$7   & \hskip 3.5pt$-$25  & \hskip 
3.5pt$-$14  & 1,2,3\\[2.8pt]
HD 104237         &  6.6 &  [T] & AHTU   & 1991.248 & 12 00 05.196  &$-$78 11 
34.51 & \hskip 3.5pt$-$43  & \hskip 7.0pt$-$5   & \hskip 3.5pt$-$41  & \hskip 
3.5pt$-$14  &  2,3 \\[2.8pt]
\hline
Lupus\\[2.8pt]
\hline
HD 139614         &  8.3 &  [T] & AETU   & 1988.539 & 15 40 46.400  &$-$42 29 
53.24 & \hskip 3.5pt$-$17  & \hskip 3.5pt$-$27  & \hskip 3.5pt$-$30  & \hskip 
3.5pt$-$11  &  1,3 \\[2.8pt]
HD 142527         &  8.4 &  [T] & AHT	 & 1991.249 & 15 56 41.899  &$-$42 19 
23.06 & \hskip 3.5pt$-$14  & \hskip 3.5pt$-$24  & \hskip 3.5pt$-$26  & \hskip 
7.0pt$-$9	& 1,2,3\\[2.8pt]
V856 Sco          &  7.0 &  [T] & AHTV   & 1991.252 & 16 08 34.294  &$-$39 06 
18.13 & \hskip 7.0pt$-$8   & \hskip 3.5pt$-$23  & \hskip 3.5pt$-$21  & \hskip 
3.5pt$-$11  &  2,3 \\[2.8pt]
\hline
U--Sco - Oph\\[2.8pt]
\hline
PDS 144 N,S       & 12.8 & [14] & AU	 & 1946.483 & 15 49 15.425  &$-$26 00 
52.48 & \hskip  14pt   0   & \hskip 7.0pt$-$8   & \hskip 7.0pt$-$5   & \hskip 
7.0pt$-$6	&      \\[2.8pt]
HD 144432         &  8.2 &  [T] & AHTU   & 1991.249 & 16 06 57.962  &$-$27 43 
09.55 & \hskip 3.5pt$-$14  & \hskip 3.5pt$-$23  & \hskip 3.5pt$-$26  & \hskip 
7.0pt$-$7	& 1,2,3\\[2.8pt]
HD 145718         &  8.9 &  [T] & AHTU   & 1991.249 & 16 13 11.597  &$-$22 29 
06.42 & \hskip 3.5pt$-$10  & \hskip 3.5pt$-$21  & \hskip 3.5pt$-$22  & \hskip 
7.0pt$-$7	& 1,2,3\\[2.8pt]
CD-23 12840       & 11.3 &  [V] & AUV	 & 1996.228 & 16 18 37.302  &$-$24 05 
22.51 & \hskip 7.0pt$-$5   & \hskip 3.5pt$-$24  & \hskip 3.5pt$-$21  & \hskip 
3.5pt$-$12  &      \\[2.8pt]
HD 147889         &  8.0 &  [T] & AHTU   & 1991.249 & 16 25 24.318  &$-$24 27 
56.34 & \hskip 7.0pt$-$4   & \hskip 3.5pt$-$26  & \hskip 3.5pt$-$22  & \hskip 
3.5pt$-$14  & 1,2,3\\[2.8pt]
HD 150193         &  8.9 &  [T] & AEHTU  & 1991.249 & 16 40 17.927  &$-$23 53 
45.03 & \hskip 7.0pt$-$8   & \hskip 3.5pt$-$18  & \hskip 3.5pt$-$19  & \hskip 
7.0pt$-$5	&  2,3 \\[2.8pt]
KK Oph            & 10.3 &  [3] & AEU	 & 1956.790 & 17 10 08.065  &$-$27 15 
18.24 & \hskip 3.5pt$-$13  & \hskip 3.5pt$-$24  & \hskip 3.5pt$-$27  & \hskip 
7.0pt$-$4	&      \\[2.8pt]
\hline
Corona Australis\\[2.8pt]
\hline
HD 176269         &  6.7 &  [T] & AHTU   & 1991.249 & 19 01 03.251  &$-$37 03 
39.06 & \hskip 3.5pt$+$11  & \hskip 3.5pt$-$29  & \hskip 3.5pt$-$23  & \hskip 
3.5pt$-$20  & 1,2,3\\[2.8pt]
HD 176270         &  6.4 &  [T] & AHTU   & 1991.249 & 19 01 04.301  &$-$37 03 
41.42 & \hskip 7.0pt$-$1   & \hskip 3.5pt$-$21  & \hskip 3.5pt$-$20  & \hskip 
7.0pt$-$6	&  2,3 \\[2.8pt]
S CrA             & 11.6 &  [9] & AEU	 & 1956.056 & 19 01 08.570  &$-$36 57 
19.41 & \hskip 7.0pt$+$9   & \hskip 3.5pt$-$34  & \hskip 3.5pt$-$28  & \hskip 
3.5pt$-$20  &      \\[2.8pt]
HD 176386         &  7.3 &  [T] & AHTU   & 1991.249 & 19 01 38.929  &$-$36 53 
26.31 & \hskip 7.0pt$+$1   & \hskip 3.5pt$-$23  & \hskip 3.5pt$-$22  & \hskip 
7.0pt$-$9	& 1,2,3\\[2.8pt]
TY CrA            &  9.5 &  [T] & AT	 & 1989.313 & 19 01 40.826  &$-$36 52 
33.55 & \hskip 7.0pt$+$2   & \hskip 3.5pt$-$31  & \hskip 3.5pt$-$29  & \hskip 
3.5pt$-$12  &   3  \\[2.8pt]
HD 177076         &  8.2 &  [T] & AHTU   & 1991.245 & 19 04 44.412  &$-$36 50 
40.72 & \hskip 7.0pt$+$4   & \hskip 3.5pt$-$23  & \hskip 3.5pt$-$20  & \hskip 
3.5pt$-$11  & 1,2,3\\[2.8pt]
\hline
Other regions\\[2.8pt]
\hline
GG Car            &  8.8 &  [T] & AHT	 & 1991.248 & 10 55 58.924  &$-$60 23 
33.47 & \hskip 7.0pt$-$7   & \hskip 7.0pt$+$5   & \hskip 7.0pt$-$8   & \hskip 
7.0pt$+$2	&  2,3 \\[2.8pt]
HD 98922          &  6.8 &  [T] & AHT	 & 1991.248 & 11 22 31.683  &$-$53 22 
11.46 & \hskip 3.5pt$-$11  & \hskip 7.0pt$+$6   & \hskip 3.5pt$-$13  & \hskip 
7.0pt$+$2	& 1,2,3\\[2.8pt]
HD 141569         &  7.1 &  [T] & AHT	 & 1991.254 & 15 49 57.759  &$-$03 55 
16.17 & \hskip 3.5pt$-$21  & \hskip 3.5pt$-$15  & \hskip 3.5pt$-$25  & \hskip 
7.0pt$+$8	& 1,2,3\\[2.8pt]
AK Sco            &  9.2 &  [T] & AHT	 & 1991.249 & 16 54 44.855  &$-$36 53 
18.31 & \hskip 3.5pt$-$14  & \hskip 3.5pt$-$27  & \hskip 3.5pt$-$30  & \hskip 
7.0pt$-$6	& 1,2,3\\[2.8pt]
51 Oph            &  4.8 &  [T] & AHTU   & 1991.249 & 17 31 24.951  &$-$23 57 
45.29 & \hskip 7.0pt$+$1   & \hskip 3.5pt$-$34  & \hskip 3.5pt$-$28  & \hskip 
3.5pt$-$19  & 1,2,3\\[2.8pt]
HD 163296         &  6.9 &  [T] & AHT	 & 1991.249 & 17 56 21.293  &$-$21 57 
21.53 & \hskip 7.0pt$-$4   & \hskip 3.5pt$-$44  & \hskip 3.5pt$-$40  & \hskip 
3.5pt$-$18  & 1,2,3\\[2.8pt]
LkH$\alpha$ 118   & 11.1 &  [3] & AE	 & 1942.841 & 18 05 49.698  &$-$24 15 
20.66 & \hskip 3.5pt$-$13  & \hskip 7.0pt$-$8   & \hskip 3.5pt$-$13  & \hskip 
7.0pt$+$7	&      \\[2.8pt]
\hline
\end{tabular}
}
See references in Table \ref{Tab6}.
\end{table*}

\begin{table*}[ht]
\scriptsize
{
\caption{\label{Tab6}Non-PMS ROSAT stars in Chamaeleon (Covino et al. 
\cite{cov}) and Lupus (Wichmann et al. \cite{wic})}
\begin{tabular}{lccccccccccc}
\hline\\[-2.8pt]
Name              &  Mag & Ref. & Sources & Epoch    & 
\multispan2{\hfill{$\alpha$\hbox{\hskip 15pt} [J2000]\hbox{\hskip 15 pt} 
$\delta$}\hfill} & 
$\mu_{\alpha}{\rm cos}\delta$ & $\mu_{\delta}$ & $\mu_{l}{\rm cos}b$ & $\mu_{b}$ 
& Other PM's\\[2.8pt] 
\omit             & \omit& \omit& \omit   &\omit     & 
\multispan2{\hfill{$\overline{{\rm{[h\hskip 6pt m\hskip 6pt s]}}\hbox{\hskip 
50pt}\hbox{\hskip 
-12pt}[^{{\rm o}}\;\;\;\arcmin\;\;\;\arcsec]}$}\hfill} & 
\multispan4{\hfill{$\overline{\hbox{\hskip 45pt}{\rm [mas/yr]}\hbox{\hskip 
45pt}}$}\hfill} & 
\omit\\[2.8pt] 
\hline\\[-7.8pt]
Chamaeleon\\[2.8pt]
\hline
RX J0849.2-7735   &  9.0 & [T] & ATU    & 1988.877 & 08 49 11.130  &$-$77 35 
58.80 & \hskip 7.0pt$-$8   & \hskip 3.5pt$+$20  & \hskip 3.5pt$-$21  & \hskip  
7.0pt$+$5   &  1,3,4,6\\[2.8pt]
RX J0853.1-8244   & 11.7 & [1] & AU     & 1940.627 & 08 53 05.137  &$-$82 43 
59.17 & \hskip 7.0pt$+$7   & \hskip 3.5pt$-$20  & \hskip 3.5pt$+$20  & \hskip  
7.0pt$-$5   &  4\\[2.8pt]
RX J0917.2-7744   & 10.7 & [T] & ATU    & 1988.664 & 09 17 10.495  &$-$77 44 
02.17 & \hskip 3.5pt$-$34  & \hskip 3.5pt$+$13  & \hskip 3.5pt$-$32  & \hskip  
3.5pt$-$17  &  1,4,6\\[2.8pt]
RX J0919.4-7738S  &  9.4 & [T] & ATU    & 1988.652 & 09 19 24.359  &$-$77 38 
45.84 &            $-$104  & \hskip 3.5pt$+$69  &             $-$120 & \hskip  
3.5pt$-$33  &  \\[2.8pt]
RX J0919.4-7738N  &  8.0 & [T] & ATU    & 1988.652 & 09 19 25.033  &$-$77 38 
37.13 &            $-$107  & \hskip 3.5pt$+$66  &             $-$120 & \hskip  
3.5pt$-$37  &  2,3,4,6\\[2.8pt]
RX J0936.3-7820   &  8.7 & [T] & AHTU   & 1991.248 & 09 36 18.035  &$-$78 20 
42.02 & \hskip 3.5pt$-$72  & \hskip 3.5pt$+$51  & \hskip 3.5pt$-$87  & \hskip  
3.5pt$-$15  &  2,3,4,6\\[2.8pt]
RX J1007.7-8504   &  8.8 & [T] & AHTU   & 1991.248 & 10 07 33.366  &$-$85 04 
36.45 &            $-$555  &            $+$390  &             $-$677 & \hskip  
3.5pt$-$52  &  2,6\\[2.8pt]
RX J1039.5-7538N  & 11.0 & [1] & AEU    & 1956.371 & 10 39 31.496  &$-$75 37 
53.27 & \hskip 3.5pt$+$10  & \hskip 7.0pt$+$0   & \hskip 7.0pt$+$8   & \hskip  
7.0pt$+$5   &  \\[2.8pt]
RX J1039.5-7538S  &  9.4 & [T] & AETU   & 1988.762 & 10 39 31.623  &$-$75 37 
56.67 & \hskip 3.5pt$+$31  & \hskip 3.5pt$+$36  & \hskip 3.5pt$+$9   & \hskip  
3.5pt$+$47  &  2,4\\[2.8pt]
RX J1120.3-7828   & 11.0 & [1] & AU     & 1941.851 & 11 20 19.444  &$-$78 28 
25.22 & \hskip 3.5pt$+$15  & \hskip 3.5pt$+$71  & \hskip 3.5pt$-$12  & \hskip  
3.5pt$+$71  &  4\\[2.8pt]
RX J1140.3-8321   & 11.6 & [1] & AU     & 1938.801 & 11 40 18.106  &$-$83 21 
02.24 & \hskip 7.0pt$-$45  & \hskip 3.5pt$+$27  & \hskip 3.5pt$-$51  & \hskip  
3.5pt$+$13  &  4\\[2.8pt]
RX J1207.9-7555   & 10.2 & [T] & AETU   & 1988.336 & 12 07 51.663  &$-$75 55 
16.06 &            $-$157  & \hskip 7.0pt$-$7   &             $-$153 & \hskip  
3.5pt$-$34  &  1,4,6\\[2.8pt]
RX J1209.8-7344   & 11.6 & [1] & AEU    & 1950.581 & 12 09 42.925  &$-$73 44 
41.28 & \hskip 3.5pt$-$12  & \hskip 7.0pt$-$4   & \hskip 3.5pt$-$11  & \hskip  
7.0pt$-$6   &  4,6\\[2.8pt]
RX J1217.4-8035   &  8.6 & [T] & ATU    & 1988.674 & 12 17 26.902  &$-$80 35 
06.73 & \hskip 7.0pt$-$1   & \hskip 3.5pt$-$11  & \hskip 7.0pt$+$1   & \hskip  
3.5pt$-$11  &  1,3,4,6\\[2.8pt]
RX J1220.6-7539   & 10.5 & [T] & AETU   & 1988.340 & 12 20 34.736  &$-$75 39 
28.78 &            $-$117  & \hskip 7.0pt$+$4   &             $-$116 & \hskip  
3.5pt$-$11  &  1,4,6\\[2.8pt]
RX J1223.5-7740   &  8.3 & [T] & AETU   & 1988.362 & 12 23 29.268  &$-$77 40 
51.51 & \hskip 3.5pt$-$63  & \hskip 3.5pt$+$10  & \hskip 3.5pt$-$64  & \hskip  
7.0pt$+$2   &  1,3,4,6\\[2.8pt]
RX J1225.3-7857   & 10.8 & [T] & ATU    & 1988.669 & 12 25 13.515  &$-$78 57 
34.52 & \hskip 3.5pt$-$24  & \hskip 3.5pt$-$23  & \hskip 3.5pt$-$21  & \hskip  
3.5pt$-$25  &  1,3,4,6\\[2.8pt]
RX J1233.5-7523   &  9.6 & [T] & AEHTU  & 1991.248 & 12 33 30.002  &$-$75 23 
11.38 & \hskip 3.5pt$-$96  & \hskip 3.5pt$+$16  & \hskip 3.5pt$-$97  & \hskip  
7.0pt$+$9   &  2,3,4,6\\[2.8pt]
RX J1325.7-7955   & 11.5 & [1] & AU     & 1938.285 & 13 25 41.842  &$-$79 55 
16.24 & \hskip 7.0pt$-$5   & \hskip 7.0pt$-$1   & \hskip 7.0pt$-$5   & \hskip 
10.5pt 0     &  4\\[2.8pt]
RX J1349.2-7549   &  9.7 & [T] & AETU   & 1988.454 & 13 49 13.108  &$-$75 49 
47.12 & \hskip 3.5pt$-$63  & \hskip 3.5pt$-$30  & \hskip 3.5pt$-$68  & \hskip  
3.5pt$-$15  &  1,3,4,6\\[2.8pt]
\hline
Lupus\\[2.8pt]
\hline
RX J1507.2-3505   & 10.8 &  [T] & AETU   & 1988.777 & 15 07 14.842  &$-$35 04 
59.26 & \hskip 3.5pt$-$32  & \hskip 3.5pt$-$28  & \hskip 3.5pt$-$42  & \hskip 
7.0pt$-$7	&   1  \\[2.8pt]
RX J1507.9-4515   & 10.8 &  [T] & ATU	 & 1988.871 & 15 07 54.472  &$-$45 15 
21.15 & \hskip 3.5pt$+$26  & \hskip 7.0pt$-$2   & \hskip 3.5pt$+$21  & \hskip 
3.5pt$-$15  &   1  \\[2.8pt]
HD 134974         & 10.4 &  [T] & AHTU   & 1991.248 & 15 14 07.559  &$-$41 03 
35.93 & \hskip 3.5pt$-$21  & \hskip 3.5pt$-$27  & \hskip 3.5pt$-$32  & \hskip 
3.5pt$-$12  &  1,2 \\[2.8pt]
RX J1518.4-3738   & 10.3 &  [T] & AETU   & 1988.765 & 15 18 26.946  &$-$37 38 
01.87 & \hskip 3.5pt$-$16  & \hskip 3.5pt$-$28  & \hskip 3.5pt$-$29  & \hskip 
3.5pt$-$14  &   1  \\[2.8pt]
RX J1524.5-3652   & 11.3 &  [7] & AEU	 & 1956.311 & 15 24 32.435  &$-$36 52 
01.66 & \hskip 3.5pt$-$21  & \hskip 3.5pt$-$24  & \hskip 3.5pt$-$31  & \hskip 
7.0pt$-$8	&      \\[2.8pt]
HD 137059         &  9.2 &  [T] & ATU	 & 1989.094 & 15 25 17.041  &$-$38 45 
25.55 & \hskip 3.5pt$-$31  & \hskip 3.5pt$-$33  & \hskip 3.5pt$-$44  & \hskip 
7.0pt$-$9	&   3  \\[2.8pt]
RX J1534.1-3916   & 10.6 &  [T] & AETU   & 1988.727 & 15 34 07.378  &$-$39 16 
17.04 & \hskip 3.5pt$-$26  & \hskip 3.5pt$-$28  & \hskip 3.5pt$-$38  & \hskip 
7.0pt$-$6	&   1  \\[2.8pt]
RX J1604.5-3207   & 11.0 &  [T] & ATU	 & 1989.023 & 16 04 30.576  &$-$32 07 
28.50 & \hskip 3.5pt$-$19  & \hskip 3.5pt$-$25  & \hskip 3.5pt$-$31  & \hskip 
7.0pt$-$5	&   1  \\[2.8pt]
HD 143978         &  9.2 &  [T] & AEHTUV & 1991.252 & 16 04 57.094  &$-$38 57 
15.29 & \hskip 3.5pt$-$30  & \hskip 3.5pt$-$48  & \hskip 3.5pt$-$54  & \hskip 
3.5pt$-$15  & 1,2,3\\[2.8pt]
\hline
\end{tabular}
}
$\bullet$References for the magnitudes (3$^{\rm rd}$ column): [1] Alcal\'a et 
al. (\cite{alb}); [2] Appenzeller et al. (\cite{app}); [3] Bastian \& Mundt 
(\cite{bas}); 
[4] Gauvin \& Strom (\cite{gau}); [5] Herbig \& Rao (\cite{hea}); [6] Hughes \& 
Hartigan (\cite{hug}); [7] Krautter et al. (\cite{kra}); [8] Lawson et al. 
(\cite{law}); 
[9] Marraco \& Rydgren (\cite{mao}); [10] Preibisch et al. (\cite{pre}) (B 
magnitudes); [11] Rydgren (\cite{rya}); [12] Rydgren (\cite{ryb}); [13] Schwartz 
\& 
Noah (\cite{scb}); 
[14] Torres (\cite{tob}); [15] Walter (\cite{waa}); [16] Walter et al. 
(\cite{wab}); [17] Wichmann et al. (\cite{wia}); [A] AC2000 (B magnitudes); [H] 
HIPPARCOS; [T] Tycho; 
[V] Valinhos.

$\bullet$Other available proper motions (last column): 1 = ACT; 2 = HIPPARCOS; 3 
= PPM (Bastian \& R\"oser \cite{bar}); 4 = Frink et al. (\cite{fri});
5 = Preibisch et al. (\cite{pre}); 6 = Terranegra et al. (\cite{ter})
\end{table*}

Out of the 213 PMS stars studied here, 101 had no previous determination of 
proper motion, as far as the consulted literature is concerned, 81
of them had known proper motions from the ACT and 41 from HIPPARCOS. In the last 
column of Tables \ref{Tab4}, \ref{Tab5} and \ref{Tab6}, 
additional references 
for other available proper motions are provided. The proper motions in common 
with the ACT and HIPPARCOS catalogues were used to externally evaluate the 
quality of our results.

A comparison of the obtained proper motions with HIPPARCOS ones is presented in
Fig. \ref{Fig2}. We can notice the good agreement
between both sets of data. The mean dispersion of the differences is 6 mas/yr, 
which is the best estimate of the external errors of our catalogue, concerning 
the bright PMS stars. For the fainter stars, no catalogues for comparison were 
found 
available.

We also compared our proper motions to the ACT ones in order to evaluate the 
importance of the addition of {\sc SERC--J} plates and meridian observations to the 
material used by ACT.
The mean dispersion of differences is 3 mas/yr.

One should stress that the depicted comparison with HIPPARCOS and ACT is not 
completely 
independent, but is useful for a first estimation of the coherence of our 
results.

\section{Discussion}

\begin{figure}
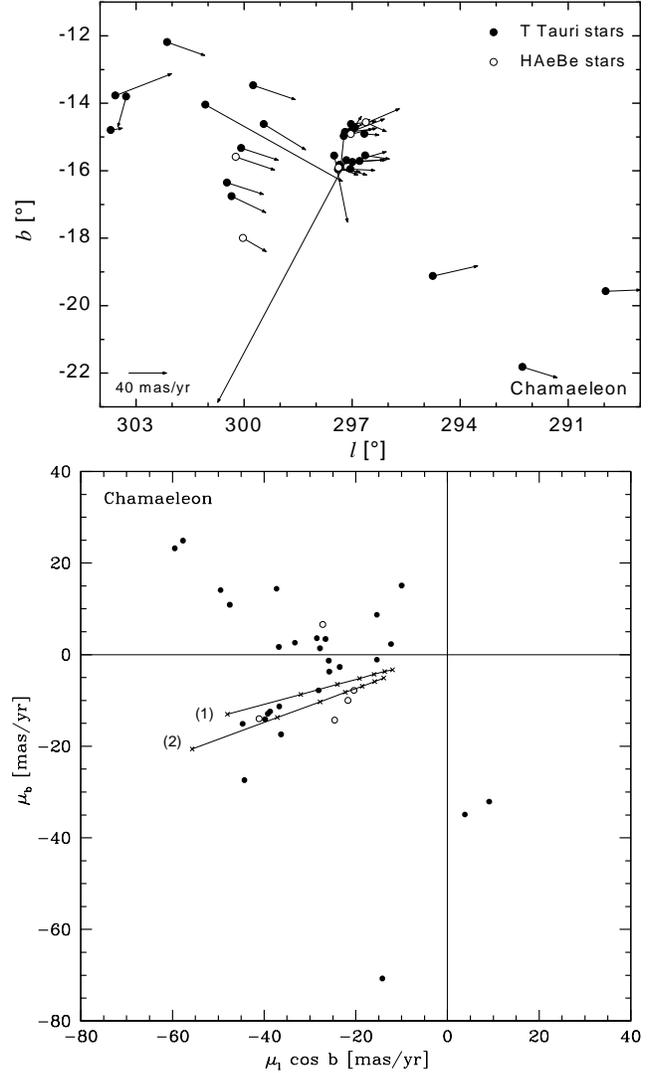

\resizebox{8.4cm}{!}{\includegraphics{ms9789.f4a}}
\resizebox{8.4cm}{!}{\includegraphics{ms9789.f4b}}
\caption{Upper panel: Positions and proper motions of PMS stars 
in Chamaeleon. Lower panel: Components of proper motions of PMS stars in 
Chamaeleon, in galactic coordinates. The reflex solar motion is presented for 
two directions -- (1): $l=289^{\rm o}$, $b=-20^{\rm o}$ and (2): $l=303^{\rm 
o}$, $b=-13^{\rm o}$ -- for distances ranging from 50 pc to 200 pc, in steps 
of 25 pc, from left to right. The open symbols represent 
HAeBe stars, the filled ones T Tauri stars.}
\label{Fig4} 
\end{figure}

\begin{figure}
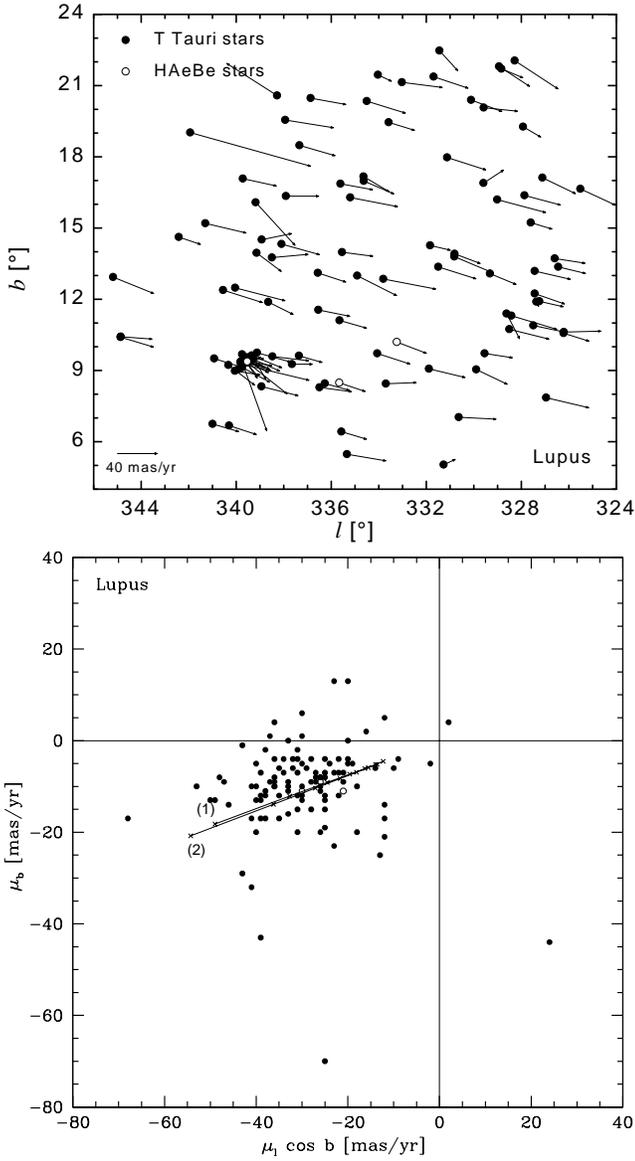

\resizebox{8.4cm}{!}{\includegraphics{ms9789.f5a}}
\resizebox{8.4cm}{!}{\includegraphics{ms9789.f5b}}
\caption{Upper panel: Positions and proper motions of PMS stars 
in Lupus. Lower panel: Same as Fig. \ref{Fig4} -- lower panel -- for Lupus. 
(1): $l=325^{\rm o}$, $b=35^{\rm o}$ and (2): $l=344^{\rm o}$, $b=10^{\rm o}$}
\label{Fig5} 
\end{figure}

\begin{figure}
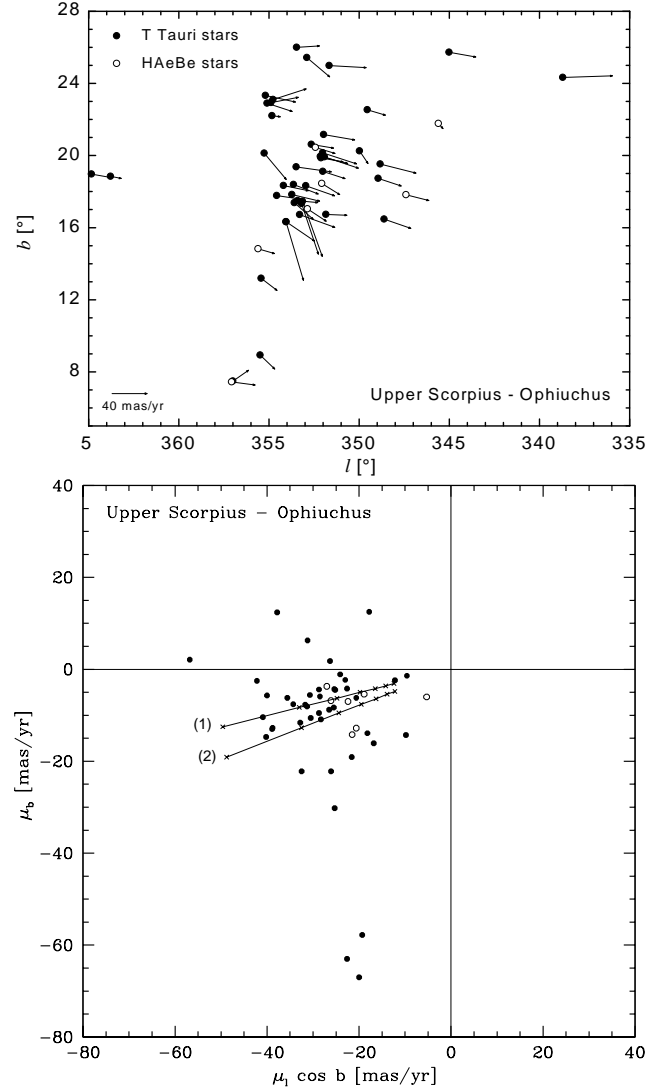

\resizebox{8.4cm}{!}{\includegraphics{ms9789.f6a}}
\resizebox{8.4cm}{!}{\includegraphics{ms9789.f6b}}
\caption{Upper panel: Positions and proper motions of PMS stars 
in Upper Scorpius -- Ophiuchus. Lower panel: Same as Fig. \ref{Fig4} -- lower 
panel -- for Upper Scorpius -- Ophiuchus.
(1): $l=345^{\rm o}$, $b=25^{\rm o}$ and (2): 
$l=355^{\rm o}$, $b=10^{\rm o}$}
\label{Fig6} 
\end{figure}

\begin{figure}
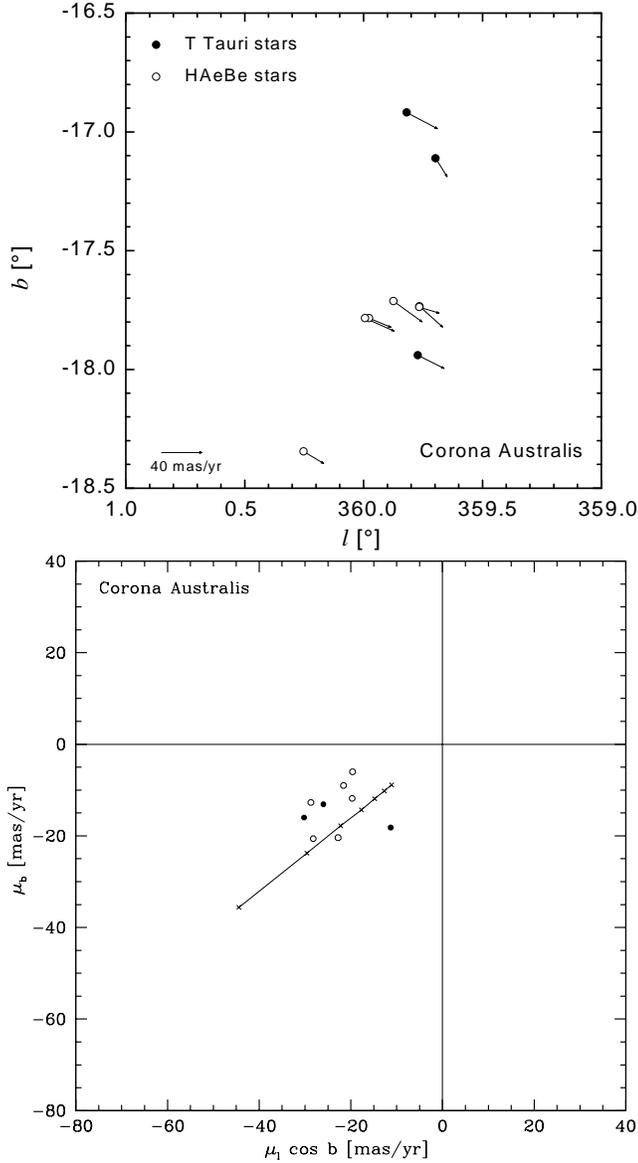

\resizebox{8.4cm}{!}{\includegraphics{ms9789.f7a}}
\resizebox{8.4cm}{!}{\includegraphics{ms9789.f7b}}
\caption{Positions and proper motions of PMS stars in Corona 
Australis. Lower panel: Same as Fig. \ref{Fig4} -- lower panel -- for Corona 
Australis. $l=359.8^{\rm o}$, $b=-17.75^{\rm o}$}
\label{Fig7} 
\end{figure}

We present in Fig. \ref{Fig3} a general view of the spatial distribution and 
proper motions
in galactic coordinates of the measured PMS stars, and in Figs. \ref{Fig4}, 
\ref{Fig5},
\ref{Fig6} and \ref{Fig7} -- upper panels -- zooms of the 4 main star-forming 
regions studied in this work. 
In all these regions, a 
dominant orientation of the proper motions towards smaller longitudes can be 
observed.
This is in large part the effect of the reflex solar motion, as can be seen in 
the
$\mu_b$ versus $\mu_l{\rm cos}b$ graphs of the same regions (Figs. \ref{Fig4},
\ref{Fig5}, \ref{Fig6} and \ref{Fig7} -- lower panels). Since the effect of 
solar motion on the star proper motion depends on the distance of the stars,
and the distances of PMS stars are 
usually poorly known, we present in these figures the reflex solar motion as 
a function 
of distance, from 50 to 200 pc. We assumed the basic solar motion, with 
components 
U = 9 km/s (directed towards the galactic center), V = 11 km/s, W = 6 km/s. The 
reflex
solar motion depends also on the direction of the stars, and since some of the 
regions
studied here have sizes of several degrees, we present it for two extreme 
directions 
in each field (except for Corona Australis, which is a small field).

In all groups, a number of stars are found to have proper motions that lie 
well
outside the average distribution. Notice that some of these stars that can be 
seen
in the ($l, b$) maps (for Upper Sco, Chamaeleon and Corona Australis) are absent 
from
the ($\mu_l{\rm cos}(b), \mu_b$) graphs, due to the scale that we adopted. We 
consider that these stars have a large probability of being recently
formed runaway stars and a possible explanation for them is the disruption of 
multiple systems. We cannot exclude the possibility of errors, like a
wrong identification during the proper motion determination process; however, 
note that some of the stars with anomalous proper motions are bright and were 
found in several sources (eg. HD 137727, HD 140637), which reduces the 
probability of an erroneous identification.

Disregarding the runaway stars, the groups of PMS stars  present proper motion 
dispersions 
of  the order of 10 mas/yr typically. This means that in a period of about one 
million 
years, the stars would move apart several degrees in the sky, and would look 
very dispersed.
We must bear in mind, however, that part of this apparent dispersion may not be 
intrinsic,
but due to the fact that the sun is approaching the group. For instance, most of 
the PMS stars
of Upper Scorpius present negative radial velocities, while most of the stars in 
Chamaeleon
present positive radial ones (e.g. Gregorio-Hetem et al. \cite{gre}; Torres et 
al. \cite{toa};
Covino et al. \cite{cov}).

Let us now comment on the average proper motion of the stellar groups. We 
already noticed 
that the average values are largely explained by the reflex solar motion. 
However,
they are not entirely due to this effect or, in other words, intrinsic average 
proper motion of the groups are detected. For instance, in the case of Lupus 
(our largest sample, Fig. \ref{Fig5}), the center of mass of the points in 
the ($\mu_l{\rm cos}(b),\mu_b$) graph is about ($-$30,$-$9), which suggests 
mean distance of about 85 pc, if we consider only the reflex solar motion. 
However, the average distance of 14 stars of this group that have parallaxes 
from 
HIPPARCOS is 138 pc. This suggests that the Lupus PMS stars have a mean 
intrinsic proper motion in the longitude direction of about $\mu_l{\rm cos}(b)$ 
= $-$10 mas/yr. 

The Chamaeleon group is peculiar, in that it seems to present two distinct 
kinematic
groups, even after excluding the runaway stars. A group with proper motions 
close to
about $\mu_l{\rm cos}(b)$ = $-$40 mas/yr, $\mu_b$ = $-$15 mas/yr, seems to be 
well behaved. Its
observed proper motion could be explained by the the reflex solar motion, if its
distance is of the order of 70 pc. And indeed, some of the stars of this group 
have
distances determined by HIPPARCOS, and are not too different from this value (T 
Cha, 66 pc, 
RX J1158.5-7754a, 86 pc, RX J1159.7-7601, 92 pc, HD 104237, 116 pc). 
This group was already discussed by Terranegra et al. (1999); notice, 
however, that Bertout et al. (\cite{ber}) consider that the HIPPARCOS distance
of T Cha is incorrect. Another group of stars
presents positive values of $\mu_b$, considerably different from the reflex 
solar motion. Among these, only HD97300 has parallax measured by HIPPARCOS (187 pc).

Finally, we remark that in the studied regions, no systematic differences 
between the proper motions
of T Tauri stars and HAeBe stars can be observed. Our results favour the PMS 
nature of the candidates
HAeBe stars included in our list.

A deeper analysis of the proper motion of the groups of PMS stars, in connection 
with 
the ages of the subgroups, will be presented in a separate paper, where our 
results
are compared with the models proposed in the literature for the mechanisms that
might have triggered the star formation.  
\begin{acknowledgements}
%---------------
The authors wish to thank Dr. M. Rapaport for helpful comments
and Dr. J. Guibert for his help in finding and measuring the SERC--J plates. 
We are
grateful to Dr. R. Neuh\"auser, for his constructive suggestions on this paper.
Special thanks are also extended to Mr. W. Monteiro, for his support concerning 
the observations. 
A partial financial support from CNPq, FAPESP, PRONEX, CAPES 
and CNRS is fully acknowledged. This work benefited from SIMBAD database and
the Astronomical Data Analysis Center, the last one being operated by the 
National Astronomical 
Observatory of Japan.

\end{acknowledgements}


\begin{thebibliography}{}
%---------------------
\bibitem[1994]{ala}
Alcal\'a J.M., 1994, Ph.D. Thesis, Ruprecht-Karls-Univ., Heidelberg
\bibitem[1995]{alb}
Alcal\'a J.M., Krautter J., Schmitt J.H.M.M., Covino E., Wichmann R., Mundt R., 
1995, A\&AS 114, 109
\bibitem[1980]{app}
Appenzeller I., Chavarria C., Krautter J., Mundt R., Wolf B., 1980, A\&A 90, 184
\bibitem[1979]{bas}
Bastian U., Mundt R., 1979, A\&AS 36, 57
\bibitem[1993]{bar}
Bastian U., R\"oser S., 1993, PPM Star Catalogue. Vols. III \& IV, 
Astronomisches Rechen-Institut, Heidelberg
\bibitem[1992]{ben}
Benevides-Soares P., Teixeira R., 1992, A\&A 253, 307
\bibitem[1999]{ber}
Bertout C., Robichon N., Arenou F., 1999, A\&A 352, 574
\bibitem[1996]{bra}
Brandner W., Alcal\'a J.M., Kunkel M., Moneti A., Zinnecker H., 1996, A\&A 307, 
121
\bibitem[1995]{cas}
Casanova S., Montmerle T., Feigelson E.D., Andr\'e P., 1995, ApJ 439, 752
\bibitem[1997]{cov}
Covino E., Alcal\'a J.M., Allain S., Bouvier J., Terranegra L., Krautter 
J., 1997, A\&A 328, 187
\bibitem[1999]{dom}
Dominici T.P., Teixeira R., Horvath J.E., Medina-Tanco G.A., Benevides-Soares 
P., 1999, A\&AS 136, 261
\bibitem[1997]{esa}
ESA, 1997, The HIPPARCOS and Tycho Catalogs, ESA SP-1200
\bibitem[1989]{fea}
Feigelson E.D., Kriss G.A., 1989, ApJ 338, 262
\bibitem[1993]{feb}
Feigelson E.D., Casanova S., Montmerle T., Guibert J., 1993, ApJ 416, 623
\bibitem[1998]{fri}
Frink S., R\"oser S., Alcal\'a J.M., Covino E., Brandner W., 1998, A\&A 338, 442
\bibitem[1992]{gau}
Gauvin L.S., Strom K.M., 1992, ApJ 385, 217
\bibitem[1992]{gre}
Gregorio-Hetem J., L\'epine J.R.D., Quast G.R., Torres C.A.O., de la Reza R., 
1992, AJ 103, 549
\bibitem[1984]{gui}
Guibert J., Charvin P., Stoclet P., 1984,  Proceedings of the 78th. Colloquium 
of the IAU, 1983, p.165 
\bibitem[1993]{har}
Hartigan P., 1993, AJ 105, 1511
\bibitem[1972]{hea}
Herbig G.H., Rao N.K., 1972, ApJ 174, 401
\bibitem[1988]{heb}
Herbig G.H., Bell K.R., 1988, Lick Observatory Bull. 1111
\bibitem[1992]{hug}
Hughes J., Hartigan P., 1992, AJ 104, 680
\bibitem[1997]{kra}
Krautter J., Wichmann R., Schmitt J.H.M.M., Alcal\'a J.M., Neuh\"auser R., 
Terranegra L., 1997, A\&AS 123, 329
\bibitem[1996]{law}
Lawson W.A., Feigelson E.D., Huenemoerder D.P., 1996, MNRAS 280, 1071
\bibitem[1998]{mal}
Malfait K., Bogaert E., Waelkens C., 1998, A\&A 331, 211
\bibitem[1981]{mao}
Marraco H.G., Rydgren A.E., 1981, AJ 86, 62
\bibitem[1998]{mat}
Mart\'{\i}n E.L., Montmerle T., Gregorio-Hetem J., Casanova S., 1998, MNRAS 300, 
733
\bibitem[1998]{mon}
Monet D., 1998 BAAS, 193, 120.03
\bibitem[1998]{neu}
Neuh\"auser R., Brandner W., 1998, A\&A 330, L29
\bibitem[1998]{pre}
Preibisch T., Guenther E., Zinnecker H., Sterzik M., Frink S., R\"oser S., 1998, 
A\&A 333, 619
\bibitem[1980a]{rya}
Rydgren A.E., 1980a, AJ 85, 438
\bibitem[1980b]{ryb}
Rydgren A.E., 1980b, AJ 85, 444
\bibitem[1977]{sca}
Schwartz R.D., 1977, ApJS 35, 161
\bibitem[1978]{scb}
Schwartz R.D., Noah P., 1978, AJ 83, 785
\bibitem[1992]{tei}
Teixeira R., R\'equi\`eme Y., Benevides-Soares P., Rapaport M., 1992, A\&A 264, 
307
\bibitem[1999]{ter}
Terranegra L., Morale F., Spagna A., Massone G., Lattanzi, M.G., 1999,
A\&A 341, L79
\bibitem[1994]{the}
Th\'e P.S., de Winter D., P\'erez M.R., 1994, A\&AS 104, 315
\bibitem[1998]{tob}
Torres C.A.O., 1998, Ph.D. Thesis, CNPq/Observat\'orio Nacional, Brazil
\bibitem[1995]{toa}
Torres C.A.O., Quast G.R., de la Reza R., Gregorio-Hetem J., L\'epine J.R.D., 
1995, AJ 109, 2146
\bibitem[1998a]{ura}
Urban S.E., Corbin T.E., Wycoff G.L., Martin J.C., Jackson E.S., Zacharias M.I., 
Hall D.M., 1998a, AJ 115, 1212
\bibitem[1998b]{urb}
Urban S.E., Corbin T.E., Wycoff G.L., 1998b, AJ 115, 2161
\bibitem[1997]{vaa}
van den Ancker M.E., Th\'e P.S., Tjin A Djie H.R.E., Catala C., de Winter D., 
Blondel P.F.C., Waters L.B.F.M., 1997, A\&A 324, L33
\bibitem[1998]{vab}
van den Ancker M.E., de Winter D., Tjin A Djie H.R.E., 1998, A\&A 330, 145
\bibitem[1999]{via}
Viateau B., R\'equi\`eme Y., Le Campion J.F., Benevides-Soares P., Teixeira 
R., Montignac G., Mazurier J.M., Monteiro W., Bosq F., Chauvet F., Colin 
J., Daigne G., Desbats J.M., Dominici T.P., P\'eri\'e J.P., Raffaelli J., 
Rapaport M., 1999, A\&AS 134, 173 
\bibitem[1986]{waa}
Walter F.M., 1986, ApJ 306, 573
\bibitem[1997]{wab}
Walter F.M., Vrba F.J., Wolk S.J., Mathieu R.D., Neuh\"auser R., 1997, AJ 114, 
1544
\bibitem[1997]{wia}
Wichmann R., Sterzik M., Krautter J., Metanomski A., Voges W., 1997, A\&A 326, 
211
\bibitem[1998]{wib}
Wichmann R., Bastian U., Krautter J., Jankovics I., Ruci\'nski S.M., 1998, MNRAS 
301, L39
\bibitem[1999]{wic}
Wichmann R., Covino E., Alcal\'a J.M., Krautter J., Allain S., Hauschildt P.H., 
1999, MNRAS 307, 909
\bibitem[1992]{wil}
Wilking B.A., Greene T.P., Lada C.J., Meyer M.R., Young E.T., 1992, ApJ 397, 520
\bibitem[1999]{zeu}
de Zeeuw P.T., Hoogerwerf R., de Bruijne H.J., Brown A.G.A., Blaauw A., 1999, AJ 
117, 354 
\end{thebibliography}
\end{document}